\documentclass[draftclsnofoot,12pt,twoside,onecolumn,letterpaper]{IEEEtran}

\usepackage[pdftex]{graphicx}
\usepackage[cmex10]{amsmath}
\usepackage{cite}
\usepackage{url}
\usepackage{color}
\usepackage{microtype}
\usepackage{amssymb}
\usepackage[font = small]{caption}
\usepackage{subcaption}
\usepackage{relsize}

\newcommand{\norm}[1]{\left\lVert#1\right\rVert}

\newcommand{\SINR}{\mbox{\small\sf\it SINR}}

\newcommand{\figlowerspace}{\vspace{-7mm}}
\begin{document}
%


\title{Sum-Rate Analysis and Optimization of Self-Backhauling Based Full-Duplex\\ Radio Access System}

%
%

\author{Dani~Korpi,
		Taneli~Riihonen,
        Ashutosh~Sabharwal,
        and~Mikko~Valkama\vspace{-15mm}
\thanks{D. Korpi and M. Valkama are with the Department of Electronics and Communications Engineering, Tampere University of Technology, Tampere, Finland (e-mail: dani.korpi@tut.fi, mikko.e.valkama@tut.fi).}%
\thanks{T. Riihonen is with the Department of Signal Processing and Acoustics, Aalto University School of Electrical Engineering, Espoo, Finland, e-mail: taneli.riihonen@aalto.fi.}
\thanks{A. Sabharwal is with the Department of Electrical and Computer Engineering, Rice University, Houston, TX 77005, USA (e-mail: ashu@rice.edu).}
\thanks{Manuscript received April 19, 2016.}}


\maketitle

\begin{abstract}
\vspace{-3mm}
In this article, a radio access system with a self-backhauling full-duplex access node serving legacy half-duplex mobile devices is studied and analyzed. In particular, it is assumed that the access node is using the same center frequency for all the transmissions, meaning that also the backhauling is done using the same frequency resources as the uplink and downlink transmissions. It is further assumed that the access node has a massive array to facilitate efficient beamforming and self-interference nulling in its own receiver. As a starting point, the signal model for the considered access node is first derived, including all the transmitted and received signals within the cell. This is then used as a basis for obtaining the sum-rate expressions, which depict the overall rates experienced by the mobile users that are served by the access node. In addition, the data rate for the bi-directional backhaul link is also derived, since the access node must be able to backhaul itself wirelessly. The maximum achievable sum-rate is then determined by numerically solving an optimization problem constructed from the data rate expressions. The full-duplex scheme is also compared to two alternative transmission schemes, which perform all or some of the transmissions in half-duplex mode. The results show that the full-duplex capability of the access node is beneficial for maximizing the sum-rate, meaning that a simple half-duplex transmission scheme is typically not optimal. In particular, the highest sum-rate is usually provided by a relay type solution, where the access node acts as a full-duplex relay between the mobiles and the backhaul node.\vspace{-3mm}
\end{abstract}


\begin{IEEEkeywords}
\vspace{-3mm}Self-backhauling, Full-duplex, Massive MIMO, Optimization, Sum-rate
\end{IEEEkeywords}

\section{Introduction}

\IEEEPARstart{W}{ireless} inband full-duplex communications is widely considered to be one key enabling technology in achieving the required throughput gains of the future 5G networks. Its basic idea is to transmit and receive data signals simultaneously using the same center-frequency, and hence it has the capability to double the spectral efficiency of the existing systems, assuming that its full potential can be harnessed properly \cite{Duarte12,Bliss07,Day12,Choi10,Jain11,Korpi14c,Goyal15a}. Many real-world demonstrations of inband full-duplex radios have already been developed by various research groups, which indicates that the concept is indeed feasible \cite{Duarte12,Choi10,Jain11,Heino15a,Bharadia13,Korpi15d}. In addition, the framework and theoretical boundaries of inband full-duplex radios have been extensively studied in the recent years \cite{Day12,Day122,Riihonen11,Riihonen13,Korpi15a,Korpi15c,Aggarwal12}.

In terms of a practical implementation, the fundamental issue for inband full-duplex devices is the coupling of the own transmit signal to the receiver. In particular, since the transmission and reception occur simultaneously over the same frequency channel, the transceiver will inherently receive its own transmit signal. What makes this especially problematic is the extremely high power level of the own transmission at this stage, which means that it will completely drown out the intended received signal. This phenomenon is typically referred to as self-interference (SI), and reducing its effect is one of the main research areas in this field. The various proposed SI cancellation solutions \cite{Anttila13,Bharadia13,Korpi14d,Ahmed13,Kaufman13,Liu14} and actual implementations and measurements already show that solving the problem of SI is not far from reality \cite{Duarte12,Heino15a,Bharadia13,Huusari15,Liu14,Korpi15d}.

In addition to SI cancellation, a large portion of the research has also focused on how to best make use of the full-duplex capability on a network level \cite{Sabharwal14,Everett11,Goyal15a}. This is a tedious issue since in many applications the traffic requirements are asymmetric between the two communicating nodes, such as in mobile networks, for instance \cite{nsn13}. Because the inband full-duplex principle assumes completely symmetric traffic to realize the doubling of spectral efficiency, this will compromise the potential throughput gains that could be achieved by it. Thus, more advanced schemes are likely needed in order to realize the full potential of inband full-duplex radios in practical network scenarios. Such schemes include the possibility of having a full-duplex access node (AN) in an otherwise legacy half-duplex mobile cell \cite{Everett11,Goyal15a}. This would allow the AN to simultaneously serve the uplink (UL) and downlink (DL) while utilizing the same frequency resources. By proper multiplexing and active scheduling, this scheme would result in the AN fully exploiting its full-duplex capability in both directions \cite{Goyal15a}.

\begin{figure*}[!t]
\centering
\includegraphics[width=0.5\textwidth]{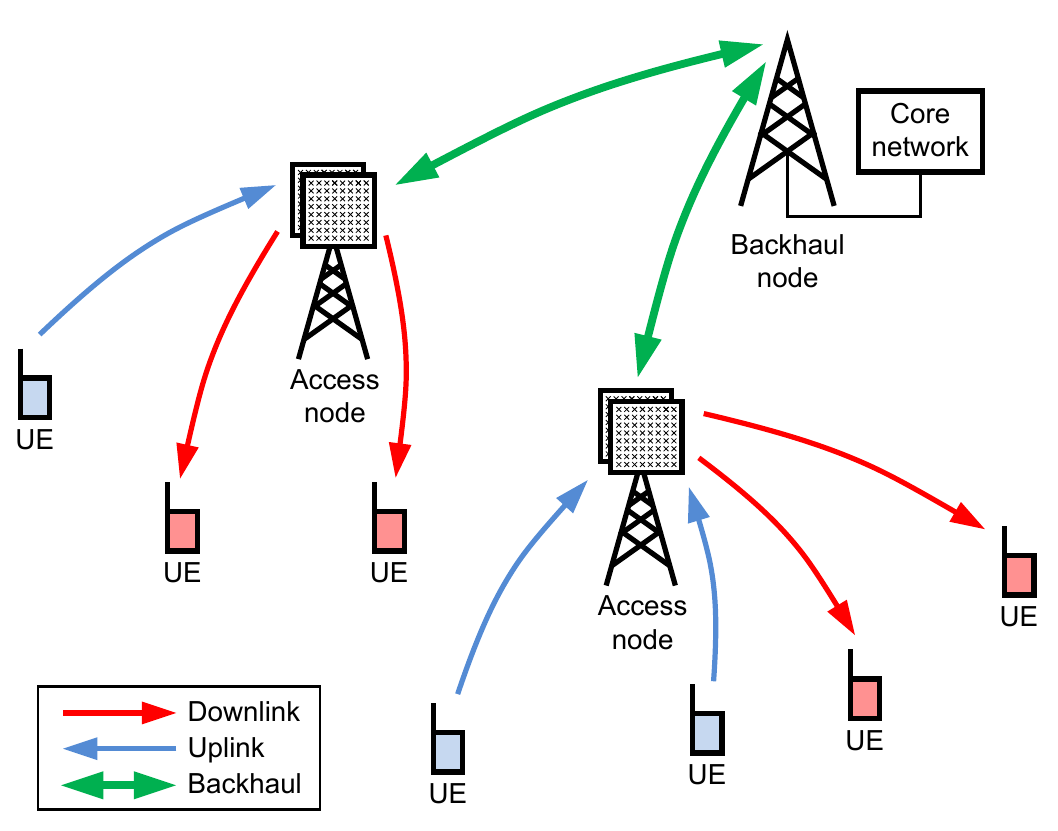}
\caption{An illustration of the considered cellular network where a full-duplex access node with large antenna arrays is serving the half-duplex UEs using wireless self-backhauling, all at the same center-frequency.}
\label{fig:syspic}
\end{figure*}

In this paper, this type of a communication scenario will be analyzed and developed further in the context of a cellular network, as illustrated in Fig.~\ref{fig:syspic}. In particular, we consider a situation where ANs are located densely to achieve high throughputs for the user equipments (UEs), for example, in a future 5G network. A special challenge in such ultra dense cellular networks is the backhaul connection, which must be able to support the mobile traffic of each cell. However, due to the dense deployment, it is very expensive to install wired backhaul links for all the cells, and thus wireless \emph{self-backhauling} is a highly lucrative solution \cite{Tabassum15a,Pitaval15a,Zhang15a,Hong14a}. In such a case, as shown in Fig.~\ref{fig:syspic}, the backhaul data is transferred with a wireless point-to-point link between the AN and some sort of a backhaul node (BN), which then further relays the data towards the actual core network using either a wired or a wireless link. To reduce the cost of deployment, it is assumed that the same BN serves several ANs.

If it is further assumed that the AN has large arrays of antennas at its disposal, the same frequency band can be used for all the transmissions. In particular, such massive antenna arrays allow for efficient beamforming, which can be used to prevent the interference between different streams \cite{Tabassum15a}. Thus, having massive transmit and receive antenna arrays, together with full-duplex capable radios, allows for using the same frequency channel for both the UL and DL, as well as for the wireless backhaul link. This results in an extremely high spectral efficiency without the additional cost of using, e.g., fiber-optic cables for the backhaul-link. In a sense, the same frequency band is reused for everything within a single cell and spatial multiplexing is used to differentiate the various streams and signals shown in Fig.~\ref{fig:syspic}. This is made possible by the massive array and the rich scattering environment of a typical mobile cell.

The massive transmit antenna array also makes it possible to attenuate SI by zero-forcing beamforming \cite{Ngo14}. Namely, the transmit signals will be precoded such that nulls are formed in the positions of all the receive antennas, which will significantly decrease the SI power coupled back to the receivers. In addition to this, more traditional SI cancellation can be performed, e.g., in the digital domain to suppress the residual SI \cite{Heino15a,Korpi14d,Korpi14c}. Nevertheless, in this article the emphasis is on the system-level analysis, and thus a baseline SI cancellation performance will be assumed in the later analysis without explicitly addressing how the SI is canceled.

The main focus in this paper is to determine the best way of utilizing the full-duplex capabilities of the considered large array AN. In particular, we will investigate whether the highest sum-rate for the mobile users will be achieved by performing all the communication at the same time, or whether there should be some type of scheduling between the different tasks. Thus, in the latter option, the different data streams visible in Fig.~\ref{fig:syspic} would also be separated in time, in addition to spatial separation. In particular, the full-duplex case is compared to a corresponding half-duplex AN, and to a relay type AN. Earlier research has already shown that an AN does not necessarily maximize its sum-rate by operating constantly in full-duplex mode \cite{Korpi15a}. Thus, this analysis provides fundamental findings regarding the most efficient communication schemes for a full-duplex capable AN in an ultra dense deployment scenario where wired backhaul connection is not an option. This is likely to be an important aspect in the context of the future 5G networks, where the ANs must be deployed much more densely to obtain the required improvements in the data rates \cite{Nokia15a,Kela15a}.

The more detailed contributions of this paper are as follows:
\begin{itemize}
\item Providing a system model for the considered cell, where the AN uses zero-forcing beamforming to both steer the beams and partially null the self-interference. Related to this, also the general signal-to-interference-plus-noise ratio (SINR) expressions for the different signal streams, and the corresponding achievable rate expressions, are derived, covering pure full-duplex, pure half-duplex and hybrid relaying system scenarios.
\item Defining the optimization problem for obtaining the highest possible sum-rate and optimal transmit powers. This is done separately for each of the considered communication schemes.
\item Providing extensive numerical results with realistic system parameter values. These results will then show which of the communication schemes obtains the highest sum-rate, and under which conditions, providing new scientific understanding related to the adoption of full-duplex radios in future networks.
\end{itemize}

The rest of this article is organized as follows. In Section~\ref{sec:system}, the system model is described and the general SINR expressions are derived. Also the different communication schemes are described in this section. The optimization problem is then defined and explained in Section~\ref{sec:optimization}, while Section~\ref{sec:results} shows the numerical results, which have been obtained by numerically solving the optimization problem. Finally, conclusions are drawn in Section~\ref{sec:conc}.


\section{System Model and Sum-rates}
\label{sec:system}

As already discussed, the leading motivation behind this work is to accommodate the large data rate demands of future wireless networks by allowing for a denser deployment of cells. This makes wireless backhaul connection an interesting aspect, as it might not be economically feasible to lay fiber optic cables for each of the cells. Hence, implementing cells or ANs capable of self-backhauling is a viable option to lower the costs of deployment \cite{Tabassum15a,Pitaval15a,Zhang15a,Hong14a}. Furthermore, by performing the backhaul data transfer on the same frequency band as the communication with the UEs, no additional spectral resources are required. This means that the frequency planning involved in the described system does not differ from that of traditional ANs.

To further improve the spectral efficiency of such cells, all the communication can be done in full-duplex mode. Assuming that the problem of SI can be managed, this can ideally result in the doubling of the data rates compared to a legacy half-duplex solution. However, it is not trivial whether doing everything in full-duplex is the optimal scheme under all circumstances. Especially, if the SI cancellation capabilities of the AN are poor, a full-duplex scheme is likely to be inferior to a scheme that avoids the SI entirely. For this reason, we will address different schemes for implementing the backhaul connection, as well as the UL and DL connections. These schemes are then analyzed and compared in terms of the achievable sum-rates.

\subsection{Generic System Model for Large Array Full-Duplex Access Node}
\label{sec:sys_model}

Let us consider a wireless cell with a large array AN that is communicating with SISO UEs and a MIMO BN. The AN is assumed to have $N_t$ transmit and $N_r$ receive antennas, and it is transmitting $M_t$ signal streams while receiving $M_r$ signal streams. Depending on the chosen operation scheme, $M_t$ and $M_r$ can consist of the data signal streams to and from the UEs and/or the backhaul data streams. Based on the assumption that the AN has a large array of antennas, we can write that $N_t >> M_t$ and $N_r >> M_r$. Now, the signals received by the UEs and/or the BN can be written as follows:
\begin{align}
	\mathbf{y} = \mathbf{H}_\mathit{t} \mathbf{x}+\mathbf{z}\text{,} \label{eq:dl_signal}
\end{align}
where $\mathbf{H}_t \in \mathbb{C}^{M_t \times N_t}$ is the total channel matrix between the AN and all the intended receivers, $\mathbf{x}\in \mathbb{C}^{N_t \times 1}$ is the transmit signal of the AN and $\mathbf{z}\in \mathbb{C}^{M_t \times 1}$ represents the different noise and interference sources. In this paper, Rayleigh fading between all communicating parties is assumed, which means that $\mathbf{H}_t \sim \mathcal{CN}(0,\mathbf{L})$, where $\mathbf{L}= \operatorname{diag}\left(L_\mathit{1}\text{, }L_\mathit{2}\text{,}\hdots\text{, }L_\mathit{M_t}\right)$ is a diagonal matrix containing the path loss normalized fading variances to the different receivers. In the continuation, to simplify the terminology and literary presentation, the path loss normalized fading variances are simply referred to as path losses.

The precoded transmit signal $\mathbf{x}$ is formed from the DL transmit data as follows:
\begin{align}
	\mathbf{x} = \mathbf{W}\mathbf{\Gamma}\mathbf{q}\text{,} \label{eq:tx_signal}
\end{align}
where $\mathbf{W} \in \mathbb{C}^{N_t \times M_t}$ is the precoding matrix, $\mathbf{\Gamma} \in \mathbb{C}^{M_t \times M_t}$ is a diagonal matrix containing the square roots of the transmit powers allocated for each symbol stream in its diagonal, and $\mathbf{q} \in \mathbb{C}^{M_t \times 1}$ contains the transmit data symbols. The power of the data symbols is assumed to be normalized as $\operatorname{E}\left[\left|{q}_\mathit{k} \right|^2\right] = 1$ where ${q}_\mathit{k}$ is the $k$th symbol. Furthermore, to allow for nulling the SI at all the receive antennas of the AN, it is assumed that $N_t > N_r+M_t$. Even though the transmitter power amplifier induced nonlinear distortion is typically a significant issue in full-duplex devices \cite{Korpi14c}, in this analysis we are using a linear signal model for simplicity. Furthermore, in a massive MIMO transmitter, the powers of the individual transmitters are typically small, alleviating the effects of the nonlinearities to some extent.  



In order to produce the desired transmit signal for each antenna from the actual transmit symbols, the precoding matrix $\mathbf{W}$ must be first defined. In this work, zero-forcing (ZF) beamforming is considered since it is usually a good solution for high SNRs \cite{Yang13}. Denoting the SI channel matrix between the AN transmit and receive antennas by $\mathbf{H}_s \in \mathbb{C}^{N_r \times N_t}$ and assuming that the AN has full channel state information (CSI) available, the ZF precoding matrix for the DL transmission in full-duplex mode can be written as \cite{Yang13}
\begin{align}
	\mathbf{W} = \mathbf{H}^H\left(\mathbf{H}\mathbf{H}^H\right)^{-1} \mathbf{\Lambda} \text{,}
\label{eq:precoder}\end{align}
where $\mathbf{H}^H = \begin{bmatrix}\mathbf{H}_t^H & \mathbf{H}_s^H\end{bmatrix}$, $(\cdot)^H$ denotes the Hermitian transpose, and $\mathbf{\Lambda} \in \mathbb{C}^{\left(M_t+N_r\right) \times M_t}$ is a non-square diagonal normalization matrix containing the individual normalization factors $\lambda_k$. The purpose of the normalization matrix is to ensure that precoding does not affect the expected effective powers of the data symbols. This is necessary in order to investigate the sum-rate with respect to the true transmit power allocated for each symbol stream, which is defined by the matrix $\mathbf{\Gamma}$. The elements of the normalization matrix are derived in Appendix~\ref{app:zf_norm} and for the full-duplex operation mode they can be expressed as
\begin{align}
	\lambda_k = \sqrt{L_\mathit{k}\left(N_t-M_t-N_r\right)} \text{.} \nonumber
\end{align}
For a half-duplex transmission period, the ZF precoding matrix is derived in a similar manner \cite{Yang13}, and also the $\lambda_k$ are otherwise identical, with the exception of $N_r$ not being subtracted since there is no need to null the receive antennas. Now, we can rewrite the received signal at the intended receivers as
\begin{align}
	\mathbf{y} = \mathbf{H}_\mathit{t} \mathbf{x}+\mathbf{z} = \mathbf{H}_\mathit{t} \mathbf{W}\mathbf{\Gamma}\mathbf{q}+\mathbf{z} = \widetilde{\mathbf{\Lambda}}\mathbf{\Gamma}\mathbf{q}+\mathbf{z} \text{,} \label{eq:final_rx_signal}
\end{align}
where $\widetilde{\mathbf{\Lambda}} \in \mathbb{C}^{M_t \times M_t}$ denotes $\mathbf{\Lambda}$ with the zero rows removed. To express the individual received signals, \eqref{eq:final_rx_signal} can alternatively be written component wise as
\begin{align}
	{y}_k = \sqrt{L_\mathit{k}\left(N_t-M_t-N_r\right)p_k}{q}_k+{z}_k \text{,} \label{eq:final_rx_signal_comp}
\end{align}
where $k = 1\text{, }2\text{,}\hdots\text{, }M_t$ and $p_k$ is the $k$th diagonal element of $\mathbf{\Gamma}$.

Stemming from \eqref{eq:final_rx_signal_comp} and assuming a large transmit antenna array, the SINR of the $k$th data signal \emph{transmitted} by the AN can then be expressed as follows for the full-duplex scheme \cite{Ngo14,Yang13}:
\begin{align}
	\SINR_\mathit{t,k} &= \frac{\operatorname{E}\left[\left|y_\mathit{k}-z_\mathit{k} \right|^2\right]}{\operatorname{E}\left[\left|z_\mathit{k} \right|^2\right]} = \frac{\operatorname{E}\left[\left|\sqrt{L_\mathit{k}\left(N_t-M_t-N_r\right)p_k }{q}_\mathit{k} \right|^2\right]}{\operatorname{E}\left[\left|{z}_\mathit{k} \right|^2\right]} = \frac{L_\mathit{k}\left(N_t-M_t-N_r\right)p_k}{\sigma_\mathit{t,N}^2+\sigma_\mathit{t,I,k}^2} \text{,} \label{eq:sinr_tx_general}
\end{align}
where $\sigma_\mathit{t,N}^2+\sigma_\mathit{t,I,k}^2$ is the variance of the noise-plus-interference term $z_k$, divided into the receiver noise and interference components. The latter includes the residual SI and inter-user interference (IUI), when present.

Using nearly an identical derivation as for the transmit signals (cf. \cite{Ngo14}), the SINR for the $j$th data signal \emph{received} by the AN can be expressed as
\begin{align}
	\SINR_\mathit{r,l} = \frac{L_\mathit{l}\left(N_r-M_r\right)p_l}{\sigma_\mathit{r,N}^2+\sigma_\mathit{r,I,l}^2} \text{,} \label{eq:sinr_rx_general}
\end{align}
where $L_\mathit{l}$ is the path loss of the $l$th signal stream, $M_r$ is the number of received signal streams, $p_l$ is the corresponding transmit power, and $\sigma_\mathit{r,N}^2+\sigma_\mathit{r,I,l}^2$ is the variance of the noise-plus-interference term. Note that, in order to make the analysis more straight-forward, the possible beamforming done by the BN is omitted from the equations. This means that, from the modeling perspective, all the processing for both the received and the transmitted backhaul signals is done by the AN. Together, the expressions in \eqref{eq:sinr_tx_general} and \eqref{eq:sinr_rx_general} can then be used to determine the transmission SINRs and the corresponding sum-rates for the different schemes, which are presented below.


\subsection{Different Communication Schemes, SINRs, and Achievable Rates}

\subsubsection{Full-duplex}
\label{sec:full-duplex}

\begin{figure*}[!t]
\centering
\includegraphics[width=0.5\textwidth]{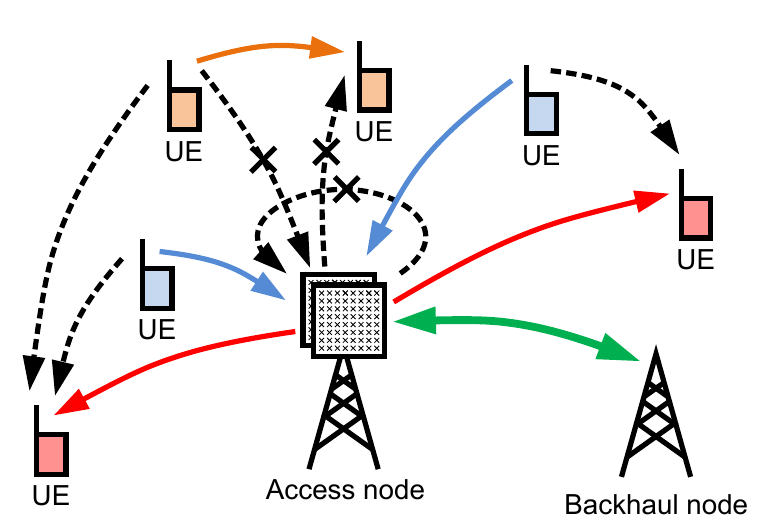}
\caption{An illustration of the full-duplex access node with wireless backhaul, where the interference signals are depicted by the dashed arrows. The impact of zero-forcing beamforming in the access node is correspondingly depicted by the cross on top of the selected interference signals.}
\label{fig:fd}
\figlowerspace
\end{figure*}

The scheme where everything is done in full-duplex mode is illustrated in Fig.~\ref{fig:fd}. The considered system consists of the actual AN, which has large antenna arrays, the BN, and the UEs. The UEs are further divided into UL and DL UEs. This type of a full-duplex system suffers from both SI and IUI. The latter is a somewhat more tedious problem than SI, as it cannot be cancelled quite as easily because of the interfering signal being now unknown. Even though there are also advanced methods for attenuating the IUI \cite{Sahai13a,Kim15a}, in this paper we assume that the only factors affecting its power level are the transmit power of the UL UEs and the path loss. In addition to serving the UEs, the AN must also maintain the bidirectional backhaul link using the same frequency band.

To consider a more general system, the forthcoming equations also account for possible intra-cell communication between the UL and DL UEs. Essentially, this means that some of the data traffic does not need to backhauled, thereby releasing resources for other tasks. Furthermore, two types of intra-cell transmissions are considered: a direct device-to-device (D2D) communication between the UL and DL UEs, and a case where the AN is relaying the traffic inside the cell. Thus, in the former, part of the IUI becomes in fact useful information, while in the latter the AN acts as a relay. As mentioned, in both of these cases the load on the backhaul link is decreased since the intra-cell traffic does not require backhauling. However, the D2D transmissions require the AN to use some degrees-of-freedom to null the interference caused by them and to ensure that it does not interfere with the UEs receiving the D2D transmissions.


As a starting point for the sum-rate analysis, the DL SINRs for this particular scheme are first determined. Assuming that each UE experiences similar fading and path loss conditions, the DL SINR for an individual UE can be approximated using \eqref{eq:sinr_tx_general} as follows:
\begin{align}
	\SINR_\mathit{d} \approx \frac{L_\mathit{UE}\left(N_t-D-M^\mathit{BH}_\mathit{t}-N_r\right) P_d}{\left(D-K_\mathit{D2D}\right)\left(\sigma_n^2+L_\mathit{UD} \left(U-K_\mathit{D2D}\right) P_u+L_\mathit{UD} K_\mathit{D2D} P_\mathit{u}^\mathit{D2D}\right)} \text{,}\label{eq:fd_sinr_dl}
\end{align}
where $L_\mathit{UE}$ is the path loss between the AN and each UE, $D$ is the total number of UEs in the DL,  $U$ is the total number of UEs in the UL, $K_\mathit{D2D}$ is the number of UE pairs communicating within the cell in a D2D fashion, $M^\mathit{BH}_\mathit{t}$ is the number of transmitted backhaul signal streams, $P_d/\left(D-K_\mathit{D2D}\right)$ is the transmit power allocated for each DL signal stream, $P_\mathit{u}^\mathit{D2D}$ is the transmit power of the D2D UEs, $\sigma_n^2$ is the receiver noise power, $L_\mathit{UD}$ is the path loss between all the UL and DL UEs, and $P_u$ is the transmit power of the UL UEs. Note that now the interference power term consists of the IUI produced by the $U$ UL UEs transmitting simultaneously to the AN and to the other D2D UEs, where the path loss between all the UL and DL UEs is assumed to be the same for simplicity. 

The UL SINR can be approximated in a similar manner, assuming that also all the UL UEs experience similar fading and path loss conditions. Using \eqref{eq:sinr_rx_general}, it can be expressed as follows:
\begin{align}
	\SINR_\mathit{u} = \frac{L_\mathit{UE}\left(N_r-U-M^\mathit{BH}_\mathit{r}\right) P_u}{\sigma_n^2+\alpha \left(P_d + P^\mathit{BH}_\mathit{u} \right)} \text{,}\label{eq:fd_sinr_ul}
\end{align}
where $\alpha$ is the amount of SI attenuation, modeling both the accuracy of the beamforming procedure as well as the effectiveness of the other SI cancellation stages, $M^\mathit{BH}_\mathit{r}$ is the number of received backhaul signal streams, and $P^\mathit{BH}_\mathit{u}$ is the total transmit power of the backhaul signal stream, transmitted by the AN.

The SINR for the D2D data streams between the UEs is of a slightly different form as it is assumed to be a SISO link. Thus, there is no beamforming gain or possibility of spatial multiplexing. The D2D SINR can simply be approximated as follows:
\begin{align}
	\SINR_\mathit{D2D} &\approx \frac{L_\mathit{UD} P_\mathit{u}^\mathit{D2D}}{\sigma_n^2+L_\mathit{UD} \left(U-K_\mathit{D2D}\right) P_u+L_\mathit{UD} \left(K_\mathit{D2D}-1\right) P_\mathit{u,D2D}}\nonumber\\
	&= \frac{1}{K_\mathit{D2D}-1+\frac{\sigma_n^2}{L_\mathit{UD} P_\mathit{u}^\mathit{D2D}}+\frac{P_\mathit{u}}{P_\mathit{u}^\mathit{D2D}}} \text{,}\label{eq:fd_sinr_D2D}
\end{align}
where it is again assumed that the path loss conditions are the same for all the UE pairs within the cell. Now, as opposed to the DL SINR, one of the streams from the UL UEs to the DL UEs is not interference, but instead constitutes the signal of interest.

The SINRs written above can then be used to calculate the different data rates for the considered system. Making the typical assumption that the noise and interference signals are Gaussian distributed, we can obtain a lower bound for the channel capacity using the Shannon-Hartley theorem \cite{Ngo14,Yang13}. In particular, for a large transmit antenna array, the achievable DL data rate can be written as follows:
\begin{align}
	C_\mathit{d} &= \sum\limits_{k \in \mathit{DL}} \operatorname{log}_2 \left(1+ \SINR_\mathit{t,k}\right) = \left(D-K_\mathit{D2D}-K_\mathit{AN}\right) \operatorname{log}_2 \left( 1+\SINR_\mathit{d} \right) \text{,} \label{eq:fd_cap_dl}
\end{align}
where $K_\mathit{AN}$ is the number of UE pairs communicating within the cell but via the AN, as opposed to the D2D UEs. In a similar fashion, for a large receive antenna array, the corresponding achievable UL data rate can be written as follows:
\begin{align}
	C_{u} &= \sum\limits_{l \in \mathit{UL}} \operatorname{log}_2 \left(1+ \SINR_\mathit{r,l}\right) = \left(U-K_\mathit{D2D}-K_\mathit{AN}\right) \operatorname{log}_2 \left( 1+\SINR_{u} \right) \label{eq:fd_cap_ul}
\end{align}
The data rate for the intra-cell traffic consists of the D2D transmissions as well as of the intra-cell traffic relayed by the AN. Noting that the data rate of the latter is defined by the SINR of the weaker link, i.e., UL or DL, the intra-cell data rate can be expressed as
\begin{align}
	C_\mathit{IC} &=\sum\limits_{m \in \mathit{D2D}} \operatorname{log}_2 \left(1+ \SINR_\mathit{D2D}\right) +\sum\limits_{n \in \mathit{IC}} \operatorname{min}\left\{\operatorname{log}_2 \left( 1+\SINR_\mathit{t,n} \right),\operatorname{log}_2 \left( 1+\SINR_\mathit{r,n} \right)\right\}\nonumber\\
	&=K_\mathit{D2D}\operatorname{log}_2 \left( 1+\SINR_\mathit{D2D} \right) + K_\mathit{AN}\operatorname{min}\left\{\operatorname{log}_2 \left( 1+\SINR_\mathit{d} \right),\operatorname{log}_2 \left( 1+\SINR_\mathit{u} \right)\right\} \text{.}\label{eq:fd_cap_D2D}
\end{align}
Using \eqref{eq:fd_cap_dl}--\eqref{eq:fd_cap_D2D}, the total sum-rate of the system can then be written as
\begin{align}
	C_\mathit{s} &= C_\mathit{d}+C_\mathit{u}+C_\mathit{IC} \nonumber\\
	&= \left(D-K_\mathit{D2D}-K_\mathit{AN}\right) \operatorname{log}_2 \left(1+\SINR_{d} \right) + \left(U-K_\mathit{D2D}-K_\mathit{AN}\right) \operatorname{log}_2 \left( 1+\SINR_\mathit{u} \right)\nonumber\\
	&+ K_\mathit{D2D}\operatorname{log}_2 \left( 1+\SINR_\mathit{D2D} \right) + K_\mathit{AN}\operatorname{min}\left\{\operatorname{log}_2 \left( 1+\SINR_\mathit{d} \right),\operatorname{log}_2 \left( 1+\SINR_\mathit{u} \right)\right\}\text{.} \label{eq:fd_cap_sum}
\end{align}
 

An important consideration for the analyzed system utilizing wireless self-backhauling is also the capacity of the backhaul connection. Namely, it must be equal to or higher than the sum-rate of the served UEs, obviously excluding the UEs engaged in communication within the cell. For this reason, the data rate of the backhaul link must be taken into account in the analysis of this type of an AN. Using a similar approach as above, the data rates of the backhaul link in the two directions can be written as follows:
\begin{align}
	C^\mathit{BH}_\mathit{d} =\sum\limits_{l \in \mathit{BH}} \operatorname{log}_2 \left(1+ \SINR_\mathit{r,l}\right) = M^\mathit{BH}_\mathit{r} \operatorname{log}_2 \left(1+\frac{L_\mathit{BH} \left(N_r-U-M^\mathit{BH}_\mathit{r}\right) P^\mathit{BH}_\mathit{d}}{M^\mathit{BH}_\mathit{r}\left[\sigma_n^2+\alpha \left(P_d + P^\mathit{BH}_\mathit{u}  \right)\right]} \right) \label{eq:fd_cap_bh_d}
\end{align}
and
\begin{align}
	C^\mathit{BH}_\mathit{u} =\sum\limits_{k \in \mathit{BH}} \operatorname{log}_2 \left(1+ \SINR_\mathit{t,k}\right) =  M^\mathit{BH}_\mathit{t} \operatorname{log}_2 \left( 1+\frac{L_\mathit{BH} \left(N_t-D-M^\mathit{BH}_\mathit{t}-N_r\right) P^\mathit{BH}_\mathit{u}}{M^\mathit{BH}_\mathit{t} \sigma_n^2} \right) \text{,} \label{eq:fd_cap_bh_u}
\end{align}
where $L_\mathit{BH}$ is the path loss of the backhaul link, and $P^\mathit{BH}_\mathit{d}$ is the total transmit power of the BN. In this work it is assumed that the BN can perfectly manage its SI by some means, while also supporting the specified numbers of spatial streams. This results in the lack of any interference terms in \eqref{eq:fd_cap_bh_u}. Assuming perfect SI cancellation in the BN is justifiable since it can be reasonably expected to be a fixed installment with a lot of computational and other hardware resources.

Now, to ensure a feasible system, the backhaul link must be able to accommodate the traffic generated by the UEs, excluding the intra-cell traffic. This means that the following constraints must be met:
\begin{align}
	C^\mathit{BH}_\mathit{d} \geq C_\mathit{d}\label{eq:fd_d_constr} \\
	C^\mathit{BH}_\mathit{u} \geq C_\mathit{u}\text{.} \label{eq:fd_u_constr}
\end{align}
As long as these constraints are fulfilled, the sum-rate given in \eqref{eq:fd_cap_sum} can be realized. This is an important aspect in the forthcoming optimization of the transmit powers, which will be discussed in more detail in Section~\ref{sec:optimization}.

\subsubsection{Half-duplex}

\begin{figure*}[!t]
\centering
\includegraphics[width=\textwidth]{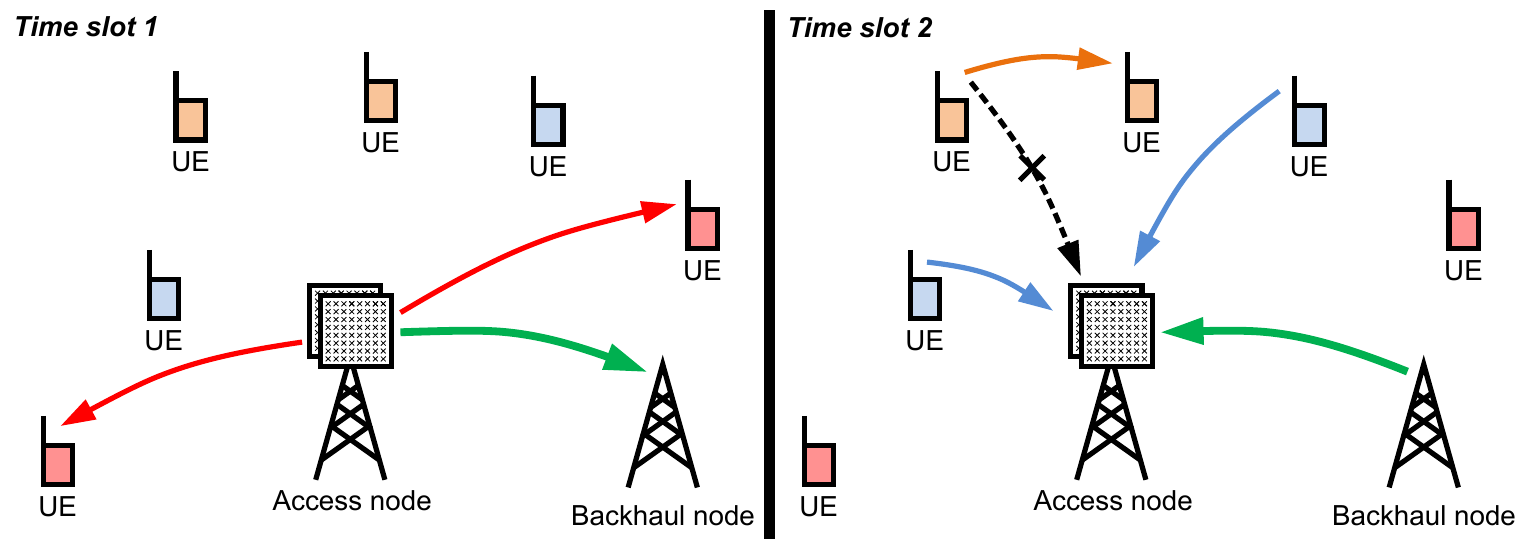}
\caption{An illustration of a half-duplex access node with wireless backhaul, where the interference signal generated by the D2D UE is depicted by the dashed arrow. The impact of zero-forcing beamforming in the access node is correspondingly depicted by the cross on top of the interference signal.}
\label{fig:hd}
\figlowerspace
\end{figure*}

One obvious alternative to the aforementioned full-duplex scheme is to introduce such scheduling that all the communication is done in half-duplex mode. In practice, such a system utilizes time division duplexing (TDD) where the whole frequency band is reserved for either transmission or reception at any given time. In terms of the analyzed AN with a wireless backhaul connection, one possible scheduling scheme is shown in Fig.~\ref{fig:hd}. There, the system has two different time slots, one for transmission and one for reception. In the former, the AN transmits data to the backhaul link and to the UEs, whereas in the latter it receives data from them. This removes the problems of SI and IUI at the cost of decreased spectral efficiency. Namely, in principle the AN now requires more temporal resources to carry out the same tasks in comparison to the full-duplex scheme. Again, it is assumed that there is a certain amount of UEs communicating only within the cell, either via the AN or directly in a D2D fashion. The D2D communication is allowed only during the UL time slot to prevent the IUI from decreasing the DL data rate. Similar to the full-duplex scheme, the AN again uses some degrees-of-freedom to null the interference caused by the D2D transmissions.

To investigate the trade-off between avoiding interference and using more temporal resources, let us derive the sum-rates also for this scheme. Following a similar derivation as in the previous section, the DL data rate can be written as follows:
\begin{align}
	C_\mathit{d} &= \eta \left(D-K_\mathit{D2D}-K_\mathit{AN}\right) \operatorname{log}_2 \left( 1+ \SINR_\mathit{d}\right)\text{,} \label{eq:hd_cap_dl}
\end{align}
where $\eta$ is the proportion of time spent in the DL time slot, and the SINR can be written as follows, based on \eqref{eq:sinr_tx_general}:
\begin{align}
\SINR_\mathit{d} = \frac{\left(N_t-D+K_\mathit{D2D}-M^\mathit{BH}_\mathit{t}\right) L_\mathit{UE} P_d}{\left(D-K_\mathit{D2D}\right) \sigma_n^2} \text{.} \label{eq:hd_sinr_dl}
\end{align}
As can be be observed from \eqref{eq:hd_sinr_dl}, now there is no interference in the DL, nor is there any loss of degrees-of-freedom due to nulling the receive antennas. However, the rate in \eqref{eq:hd_cap_dl} is decreased by the time division duplexing factor $\eta$ since the available time must be divided between transmission and reception. The corresponding UL data rate can then be expressed as follows:
\begin{align}
	C_\mathit{u} &= \left(1 - \eta\right) \left(U-K_\mathit{D2D}-K_\mathit{AN}\right) \operatorname{log}_2 \left( 1+\SINR_\mathit{u}\right)\text{,} \label{eq:hd_cap_ul}
\end{align}
where
\begin{align}
\SINR_\mathit{u} = \frac{\left(N_r-U-M^\mathit{BH}_\mathit{r}\right) L_\mathit{UE} P_u}{\sigma_n^2}\text{.} \label{eq:hd_sinr_ul}
\end{align}

The data rate of the intra-cell traffic can be defined in a similar form as in the full-duplex scheme, the only difference being that now the direct D2D transmissions are allowed only during the UL time slot.  Hence, the overall data rate for the intra-cell traffic can be written as
\begin{align}
	 C_\mathit{IC} &= \left(1 - \eta\right) K_\mathit{D2D}  \operatorname{log}_2 \left( 1+\SINR_\mathit{D2D} \right)\nonumber\\
	&+ K_\mathit{AN} \operatorname{min} \left\{\eta\operatorname{log}_2 \left( 1+ \SINR_\mathit{d}\right),\left(1-\eta\right)\operatorname{log}_2 \left( 1+ \SINR_\mathit{u}\right) \right\} \label{eq:hd_cap_D2D}
\end{align}
where $\SINR_\mathit{D2D}$ is as defined in \eqref{eq:fd_sinr_D2D}. Note that the D2D transmissions still suffer from IUI since they occur simultaneously with all the other UL transmissions.

Based on \eqref{eq:hd_cap_dl}, \eqref{eq:hd_cap_ul}, and \eqref{eq:hd_cap_D2D}, the sum-rate of the half-duplex scheme can finally be written as follows:
\begin{align}
	C_\mathit{s} &= \eta \left(D-K_\mathit{D2D}-K_\mathit{AN}\right) \operatorname{log}_2 \left( 1+ \SINR_\mathit{d}\right)\nonumber\\
	&+\left(1 - \eta\right) \left(U-K_\mathit{D2D}-K_\mathit{AN}\right) \operatorname{log}_2 \left( 1+\SINR_\mathit{u}\right)\nonumber\\
	&+\left(1 - \eta\right) K_\mathit{D2D} \operatorname{log}_2 \left( 1+\SINR_\mathit{D2D} \right) \nonumber\\
	&+ K_\mathit{AN} \operatorname{min} \left\{\eta\operatorname{log}_2 \left( 1+ \SINR_\mathit{d}\right),\left(1-\eta\right)\operatorname{log}_2 \left( 1+ \SINR_\mathit{u}\right) \right\}\text{.} \label{eq:hd_cap_sum}
\end{align}

Again, the backhaul link must be able to provide a sufficient data rate to enable the AN to serve all the UEs. Similar to the full-duplex case, the transmission and reception data rates of the backhaul link can be written as
\begin{align}
	C^\mathit{BH}_\mathit{u} = \eta M^\mathit{BH}_\mathit{t} \operatorname{log}_2 \left( 1+\frac{\left(N_t-D+K_\mathit{D2D}-M^\mathit{BH}_\mathit{t}\right) L_\mathit{BH} P^\mathit{BH}_\mathit{u}}{M^\mathit{BH}_\mathit{t} \sigma_n^2} \right) \label{eq:hd_cap_bh_u}
\end{align}
and
\begin{align}
	C^\mathit{BH}_\mathit{d} = \left(1 - \eta\right) M^\mathit{BH}_\mathit{r} \operatorname{log}_2 \left( 1+\frac{\left(N_r-U-M^\mathit{BH}_\mathit{r}\right) L_\mathit{BH} P^\mathit{BH}_\mathit{d}}{M^\mathit{BH}_\mathit{r} \sigma_n^2} \right) \text{,}\label{eq:hd_cap_bh_d}
\end{align}
As illustrated also in Fig.~\ref{fig:hd}, now the AN does its backhaul transmission in the same time slot as the DL transmission to avoid simultaneous transmission and reception. The relationship between the sum-rate of the AN and the data rate of the backhaul link is again the same as specified in \eqref{eq:fd_d_constr} and \eqref{eq:fd_u_constr}.

Based on \eqref{eq:hd_cap_dl}--\eqref{eq:hd_cap_bh_d}, an intuitive interpretation regarding the relationship between the sum-rates of the full-duplex and half-duplex schemes is that it highly depends on the level of the total interference, consisting of both the SI and the IUI. With low path loss between the UL and DL UEs and poor SI cancellation performance, the half-duplex scheme is likely to be the better option due to it being immune to the interference. On the other hand, if the AN is capable of efficiently suppressing the SI signal, and the UEs do not strongly interfere with each other, the full-duplex scheme will probably provide the higher sum-rate. This will be demonstrated through the explicit sum-rate optimization in Sections~\ref{sec:optimization}--\ref{sec:results}.

\subsubsection{Hybrid Relay}

\begin{figure*}[!t]
\centering
\includegraphics[width=\textwidth]{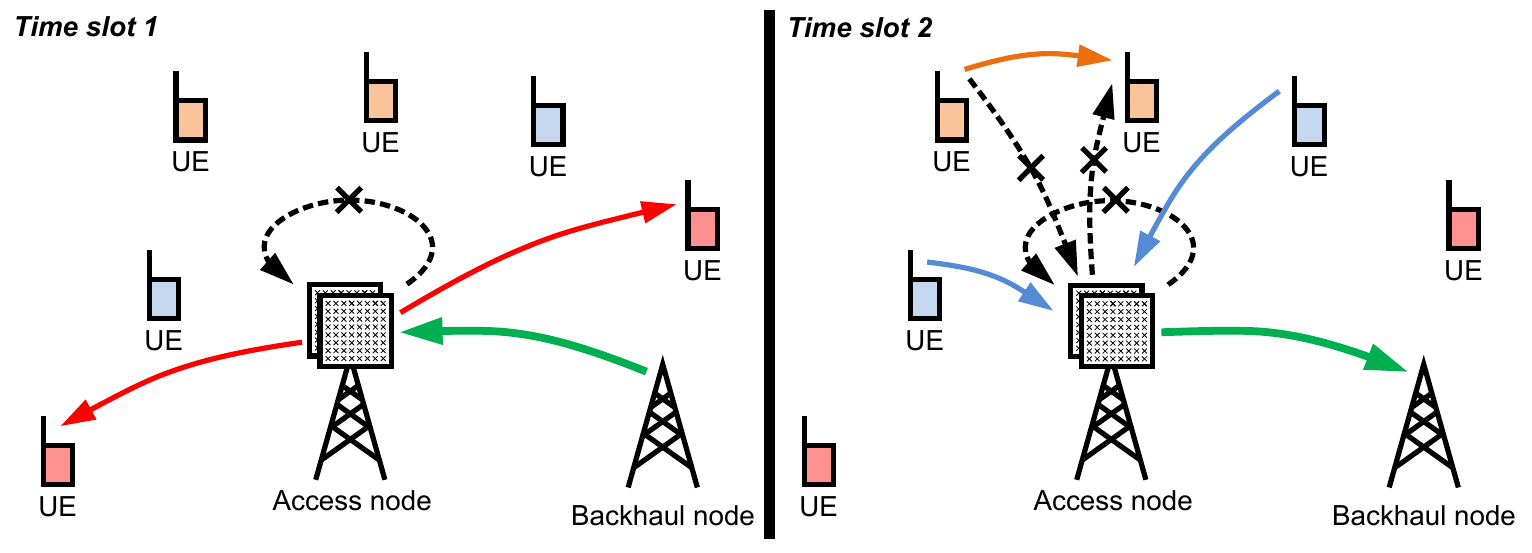}
\caption{An illustration of a relay type access node with wireless backhaul, where the interference signals are depicted by the dashed arrows. The impact of zero-forcing beamforming in the access node is correspondingly depicted by the cross on top of the selected interference signals.}
\label{fig:relay}
\figlowerspace
\end{figure*}

In addition to the above extreme cases of purely full-duplex and half-duplex systems, an interesting scheme is a relay type AN, which simply relays the UL signal to the BN during one time slot, and then in the other time slot relays the signal from the BN to the DL UEs. Figure~\ref{fig:relay} illustrates this type of a solution. The benefit of this scheme is that the problem of IUI is completely avoided, similar to the half-duplex scheme, while the full-duplex capability of the AN is still utilized to some extent. The obvious drawback is, however, that now everything cannot be done simultaneously, which will inherently decrease the achievable rate.

The DL and UL data rates for this scheme are derived similar to the other schemes, and they can be expressed as follows:
\begin{align}
	C_\mathit{d} &= \eta \left(D-K_\mathit{D2D}-K_\mathit{AN}\right) \operatorname{log}_2 \left( 1+\SINR_\mathit{d} \right)\label{eq:relay_cap_dl}\\
	C_\mathit{u} &= \left(1 - \eta\right) \left(U-K_\mathit{D2D}-K_\mathit{AN}\right) \operatorname{log}_2 \left( 1+\SINR_\mathit{u} \right) \text{,} \label{eq:relay_cap_ul}
\end{align}
where the SINRs, again given directly by \eqref{eq:sinr_tx_general} and \eqref{eq:sinr_rx_general}, are
\begin{align}
\SINR_\mathit{d} = \frac{\left(N_t-D+K_\mathit{D2D}-N_r\right) L_\mathit{UE} P_d}{\left(D-K_\mathit{D2D}\right) \sigma_n^2} \label{eq:rl_sinr_dl}
\end{align}
and
\begin{align}
	\SINR_\mathit{u} = \frac{\left(N_r-U\right) L_\mathit{UE} P_u}{\sigma_n^2 + \alpha P^\mathit{BH}_\mathit{u}} \text{.} \label{eq:rl_sinr_ul}
\end{align}
Since the backhaul-transmission occurs now in the UL time slot, the UL signal suffers from SI, which will result in a somewhat decreased data rate.

The data rate of the intra-cell traffic is of the same form as in the half-duplex scheme, and it is given directly by \eqref{eq:hd_cap_D2D}. Also the SINR of the D2D transmissions is the same as in the half-duplex and full-duplex schemes, and thus only the DL and UL SINRs must be replaced by \eqref{eq:rl_sinr_dl} and \eqref{eq:rl_sinr_ul}, respectively. Note that the D2D transmissions are again performed during the UL time slot to prevent them from interfering with the DL signals. Thus, also in this scheme, only the D2D transmissions suffer from IUI.

The expressions in \eqref{eq:hd_cap_D2D}, \eqref{eq:relay_cap_dl}, and \eqref{eq:relay_cap_ul} then allow us to define the sum-rate of the hybrid relay scheme, which can be expressed as follows:
\begin{align}
	C_\mathit{s} &= \eta \left(D-K_\mathit{D2D}-K_\mathit{AN}\right) \operatorname{log}_2 \left( 1+ \SINR_\mathit{d}\right)+\left(1 - \eta\right) \left(U-K_\mathit{D2D}-K_\mathit{AN}\right) \operatorname{log}_2 \left( 1+\SINR_\mathit{u}\right)\nonumber\\
	&+\left(1 - \eta\right) K_\mathit{D2D} \operatorname{log}_2 \left( 1+\SINR_\mathit{D2D} \right)\nonumber\\
	&+K_\mathit{AN} \operatorname{min} \left\{\eta\operatorname{log}_2 \left( 1+ \SINR_\mathit{d}\right),\left(1-\eta\right)\operatorname{log}_2 \left( 1+ \SINR_\mathit{u}\right) \right\}\text{.} \label{eq:relay_cap_sum}
\end{align}
This is of the same form as the sum-rate of the half-duplex scheme, the only difference being that the UL and DL SINRs are different for the two schemes.

In this scheme, the data rate of the incoming backhaul signal also suffers from SI due to the simultaneous DL transmission. Hence, the data rates for the backhaul link can now be written as follows:
\begin{align}
	C^\mathit{BH}_\mathit{d} = \eta M^\mathit{BH}_\mathit{r} \operatorname{log}_2 \left( 1+\frac{\left(N_r-M^\mathit{BH}_\mathit{r}\right) L_\mathit{BH} P^\mathit{BH}_\mathit{d}}{M^\mathit{BH}_\mathit{r}\left(\sigma_n^2 + \alpha P_d\right)} \right) \text{,}\label{eq:relay_cap_bh_d}
\end{align}
\begin{align}
	C^\mathit{BH}_\mathit{u} = \left(1 - \eta\right) M^\mathit{BH}_\mathit{t} \operatorname{log}_2 \left( 1+\frac{\left(N_t-M^\mathit{BH}_\mathit{t}-K_\mathit{D2D}-N_r\right) L_\mathit{BH} P^\mathit{BH}_\mathit{u}}{M^\mathit{BH}_\mathit{t} \sigma_n^2} \right) \text{.} \label{eq:relay_cap_bh_u}
\end{align}
Due to the simultaneous transmission periods for the backhaul and D2D signals, the AN must form nulls to the UEs receiving the D2D transmissions to avoid interfering with them, similar to the other schemes. Again, a crucial aspect of the considered cell is that the backhaul link should be able to match the data rates of UL and DL. Otherwise the system will be bottlenecked by the backhaul connection.

\section{Optimizing the Sum-rates}
\label{sec:optimization}

Let us next define the optimization problem that maximizes the sum-rate of the self-backhauling AN under the different considered communication schemes. The maximization can be done based on the rate expressions derived in Section~\ref{sec:system}, taking into account the necessary constraints, which ensure the feasibility of the system.

\subsection{Constraints}

A fundamental system level constraint is given by \eqref{eq:fd_d_constr} and \eqref{eq:fd_u_constr}, and it applies to all the different schemes. The constraint specifies that the backhaul connection must have an equal or higher rate than the UL and DL connections. In other words, the AN should balance its resources such that the backhaul connection is not limiting the overall rate, while maximizing the rate experienced by the mobile users. This gives us the following inequality constraints:
\begin{align}
	g_1 \left(\boldsymbol{\lambda}\right) = C_{d} - C^\mathit{BH}_\mathit{d} \leq 0 \label{eq:bh_d_constr} \\
	g_2 \left(\boldsymbol{\lambda}\right) = C_{u} - C^\mathit{BH}_\mathit{u} \leq 0  \text{,} \label{eq:bh_u_constr}
\end{align}
where $\mathbf{\lambda}$ includes the parameters to be optimized. Note that the backhaul link only has to provide enough data rate for the outgoing and incoming traffic, not for the traffic occurring within the cell. Hence, the term $C_\mathit{IC}$ is not included into the constraints in \eqref{eq:bh_d_constr} and \eqref{eq:bh_u_constr}.

To obtain meaningful results, the transmit powers must also be restricted. This means that the available transmit powers of the AN, the BN, and the UEs are upper-limited. For the AN, this gives the following constraint:
\begin{align}
	g_3 \left(\boldsymbol{\lambda}\right) = P_{d}+P^\mathit{BH}_\mathit{u} - P_{AN} \leq 0 \text{,} \label{eq:bs_pwr_const}
\end{align}
where $P_{AN}$ is the predefined maximum transmit power of the AN. In the hybrid relay scheme, \eqref{eq:bs_pwr_const} is reduced to a simple inequality requirement for a single transmit power, as $P_{d}$ and $P^\mathit{BH}_\mathit{u}$ cannot both be non-zero at the same time.

Similar restrictions are applied also to the other transmitting parties. For them, the constraints are expressed as follows:
\begin{align}
	g_4 \left(\boldsymbol{\lambda}\right) = P_{u} - P_\mathit{UE} \leq 0 \label{eq:ue_pwr_const}
\end{align}
\begin{align}
	g_5 \left(\boldsymbol{\lambda}\right) = P_\mathit{u,D2D} - P_\mathit{UE} \leq 0 \label{eq:ue_D2D_pwr_const}
\end{align}
\begin{align}
  g_6 \left(\boldsymbol{\lambda}\right) = P^\mathit{BH}_\mathit{d} - P^\mathit{BH}_\mathit{d,max} \leq 0 \text{,}\label{eq:bh_pwr_const}
\end{align}
where $P_\mathit{UE}$ is the maximum transmit power of a single UE, and $P^\mathit{BH}_\mathit{d,max}$ is the maximum transmit power of the BN.



An important consideration is also the relationship between the UL and DL data rates. In particular, it has been shown that typically the data rate requirements are heavily asymmetrical between UL and DL, such that the required UL data rate is only a fraction of the required DL data rate \cite{Falaki10}. In this analysis, the asymmetry is taken into account by establishing a relationship between the data rates, which is expressed as
\begin{align}
	C_u = \rho C_d\text{,} \label{eq:rate_ratio}
\end{align}
where $\rho$ is the ratio between the UL and DL data rates. This requirement ensures that the total sum-rate is divided in a proper manner between the UL and DL users. To allow for sufficient flexibility in the data rate requirements, in the actual optimization $\rho$ is allowed to vary within certain boundaries, i.e., $\rho_\mathit{min} \leq \rho \leq \rho_\mathit{max}$. This gives us the following inequality constraints:
\begin{align}
	g_7 \left(\boldsymbol{\lambda}\right) = \rho_\mathit{min} C_d - C_u \leq 0 \label{eq:rho_min_constr}
\end{align}
\begin{align}
	g_8 \left(\boldsymbol{\lambda}\right) =  C_u - \rho_\mathit{max} C_d \leq 0 \text{.}\label{eq:rho_max_constr}
\end{align}
Note that the intra-cell traffic, represented by $C_\mathit{IC}$, is left out of the rate ratio constraints, meaning that it is treated as an independent and unrestricted component in the sum-rate.

In addition, the half-duplex and hybrid relay schemes involve the parameter $\eta$, which specifies the proportional lengths of the two time slots. It is constrained by
\begin{align}
	0 \leq \eta \leq 1 \text{.} \label{eq:dl_eta_const}
\end{align}
This obviously stems from $\eta$ being essentially the percentage of time spent in the time slot during which DL transmissions are conducted. Equation \eqref{eq:dl_eta_const} can be transformed into two inequality constraints as follows:
\begin{align}
	g_7 \left(\boldsymbol{\lambda}\right) = -\eta \leq 0 \label{eq:eta_min_constr}
\end{align}
\begin{align}
	g_8 \left(\boldsymbol{\lambda}\right) =  \eta - 1 \leq 0 \text{.}\label{eq:eta_max_constr}
\end{align}
Together, \eqref{eq:bh_d_constr}--\eqref{eq:eta_max_constr} provide the constraints for the optimization problem, which aims at maximizing the total sum-rate experienced by the mobile users. Below, the actual optimization problem will be defined and discussed in detail.

\subsection{Optimization Problem}

The actual maximization of the sum-rate is done by selecting the optimal transmit powers for all communicating parties. In general, the optimization problem for maximizing the sum-rate can be written as follows:
\begin{gather}
	S^\mathit{X}_\mathit{max} = \operatorname*{max}_{\boldsymbol{\lambda}} S^\mathit{X}\left(\boldsymbol{\lambda}\right) \text{,} \label{eq:opt_max}\\
	\text{s.t. } \mathbf{g} \left(\boldsymbol{\lambda}\right) \leq \mathbf{0}\nonumber
\end{gather}
Here, $X \in \left\{\mathit{FD, HD, RL}\right\}$ and $S^\mathit{X}\left(\boldsymbol{\lambda}\right)$ is the sum-rate with respect to the parameter vector defined as
\begin{align}
	&\boldsymbol{\lambda} = \begin{bmatrix}P_d & P_u & P_\mathit{d}^\mathit{BH} & P_\mathit{u}^\mathit{BH} & P_\mathit{u}^\mathit{D2D} & \eta\end{bmatrix}
\end{align}
and $\mathbf{g}$ is a vector-valued function containing the constraints as
\begin{align}
	\mathbf{g} \left(\boldsymbol{\lambda}\right) = \begin{bmatrix}g_1\left(\boldsymbol{\lambda}\right) & g_2\left(\boldsymbol{\lambda}\right) & \cdots & g_8\left(\boldsymbol{\lambda}\right) \end{bmatrix}^T \text{.} \label{eq:g_all}
\end{align}
Note that in the full-duplex scheme the time division parameter $\eta$ does not affect the sum-rate and thereby its value can be set arbitrarily in that case. Furthermore, in the cases where there are no D2D transmissions, i.e., $K_\mathit{D2D} = 0$, the value for $P_\mathit{u}^\mathit{D2D}$ is set to zero.

Now, we can finally write the optimization functions for all the three schemes. For the full-duplex scheme, it can be expressed as
{\small\begin{align}
S^\mathit{FD}\left(\boldsymbol{\lambda}\right) &= \left(D-K_\mathit{D2D}-K_\mathit{AN}\right)\operatorname{log}_2 \left( 1+\frac{L_\mathit{UE}\left(N_t-D-M^\mathit{BH}_\mathit{t}-N_r\right) P_d}{\left(D-K_\mathit{D2D}\right)\left(\sigma_n^2+L_\mathit{UD} \left(U-K_\mathit{D2D}\right) P_u+L_\mathit{UD} K_\mathit{D2D} P_\mathit{u}^\mathit{D2D}\right)}  \right)\nonumber\\
&+\left(U-K_\mathit{D2D}-K_\mathit{AN}\right) \operatorname{log}_2 \left( 1+\frac{L_\mathit{UE}\left(N_r-U-M^\mathit{BH}_\mathit{r}\right) P_u}{\sigma_n^2+\alpha \left(P_d + P^\mathit{BH}_\mathit{u} \right)} \right)\nonumber\\
&+K_\mathit{D2D}\operatorname{log}_2 \left( 1+\frac{1}{K_\mathit{D2D}-1+\frac{\sigma_n^2}{L_\mathit{UD} P_\mathit{u}^\mathit{D2D}}+\frac{P_\mathit{u}}{P_\mathit{u}^\mathit{D2D}}} \right)\nonumber\\ 
&+ K_\mathit{AN}\operatorname{min}\left\{\operatorname{log}_2 \left( 1+\frac{L_\mathit{UE}\left(N_t-D-M^\mathit{BH}_\mathit{t}-N_r\right) P_d}{\left(D-K_\mathit{D2D}\right)\left(\sigma_n^2+L_\mathit{UD} \left(U-K_\mathit{D2D}\right) P_u+L_\mathit{UD} K_\mathit{D2D} P_\mathit{u}^\mathit{D2D}\right)} \right),\right.\nonumber\\
&\hspace{64.5mm}\left.\operatorname{log}_2 \left( 1+\frac{L_\mathit{UE}\left(N_r-U-M^\mathit{BH}_\mathit{r}\right) P_u}{\sigma_n^2+\alpha \left(P_d + P^\mathit{BH}_\mathit{u} \right)} \right)\right\} \text{.} \label{eq:opt_fd}
\end{align}}
For the half-duplex scheme, the sum-rate can be written as
\begin{align}
S^\mathit{HD}\left(\boldsymbol{\lambda}\right) &= \eta\left(D-K_\mathit{D2D}-K_\mathit{AN}\right) \operatorname{log}_2 \left( 1+\frac{\left(N_t-D+K_\mathit{D2D}-M^\mathit{BH}_\mathit{t}\right) L_\mathit{UE} P_d}{\left(D-K_\mathit{D2D}\right) \sigma_n^2}  \right)\nonumber\\
&+\left(1-\eta\right)\left(U-K_\mathit{D2D}-K_\mathit{AN}\right) \operatorname{log}_2 \left( 1+ \frac{\left(N_r-U-M^\mathit{BH}_\mathit{r}\right) L_\mathit{UE} P_u}{\sigma_n^2} \right)\nonumber\\
&+\left(1-\eta\right) K_\mathit{D2D}\operatorname{log}_2 \left( 1+\frac{1}{K_\mathit{D2D}-1+\frac{\sigma_n^2}{L_\mathit{UD} P_\mathit{u}^\mathit{D2D}}+\frac{P_\mathit{u}}{P_\mathit{u}^\mathit{D2D}}} \right)\nonumber\\ 
&+ K_\mathit{AN}\operatorname{min}\left\{\eta\operatorname{log}_2 \left( 1+\frac{\left(N_t-D+K_\mathit{D2D}-M^\mathit{BH}_\mathit{t}\right) L_\mathit{UE} P_d}{\left(D-K_\mathit{D2D}\right) \sigma_n^2} \right),\right.\nonumber\\
&\hspace{20mm}\left.\left(1-\eta\right)\operatorname{log}_2 \left( 1+\frac{\left(N_r-U-M^\mathit{BH}_\mathit{r}\right) L_\mathit{UE} P_u}{\sigma_n^2} \right)\right\} \text{.} \label{eq:opt_hd}
\end{align}
Finally, the sum-rate of the hybrid relay scheme is as follows:
\begin{align}
S^\mathit{RL}\left(\boldsymbol{\lambda}\right) &= \eta\left(D-K_\mathit{D2D}-K_\mathit{AN}\right) \operatorname{log}_2 \left( 1+\frac{\left(N_t-D+K_\mathit{D2D}-N_r\right) L_\mathit{UE} P_d}{\left(D-K_\mathit{D2D}\right) \sigma_n^2} \right)\nonumber\\
&+\left(1-\eta\right)\left(U-K_\mathit{D2D}-K_\mathit{AN}\right) \operatorname{log}_2 \left( 1+ \frac{\left(N_r-U\right) L_\mathit{UE} P_u}{\sigma_n^2 + \alpha P^\mathit{BH}_\mathit{u}} \right)\nonumber\\
&+\left(1-\eta\right) K_\mathit{D2D}\operatorname{log}_2 \left( 1+\frac{1}{K_\mathit{D2D}-1+\frac{\sigma_n^2}{L_\mathit{UD} P_\mathit{u}^\mathit{D2D}}+\frac{P_\mathit{u}}{P_\mathit{u}^\mathit{D2D}}} \right)\nonumber\\ 
&+ K_\mathit{AN}\operatorname{min}\left\{\eta\operatorname{log}_2 \left( 1+\frac{\left(N_t-D+K_\mathit{D2D}-N_r\right) L_\mathit{UE} P_d}{\left(D-K_\mathit{D2D}\right) \sigma_n^2} \right),\right.\nonumber\\
&\hspace{20mm}\left.\left(1-\eta\right)\operatorname{log}_2 \left( 1+\frac{\left(N_r-U\right) L_\mathit{UE} P_u}{\sigma_n^2 + \alpha P^\mathit{BH}_\mathit{u}} \right)\right\} \text{.} \label{eq:opt_rl}
\end{align}
Equations \eqref{eq:opt_fd}, \eqref{eq:opt_hd}, and \eqref{eq:opt_rl} are then used as described in \eqref{eq:opt_max} to determine the optimal sum-rates for the three considered communication schemes under different circumstances.

\section{Numerical Results}
\label{sec:results}

The numerical values for the optimal sum-rates are evaluated using the example parameter values specified in Table~\ref{table:def_param}. These values are used in the calculations unless otherwise stated. The optimization is performed with Matlab using the function \texttt{fmincon}, which is capable of performing optimization under nonlinear inequality constraints. The optimization routine is then run with several random initial guesses to ensure that the true global maximum is found with a reasonable certainty. The following results show then the optimized sum-rates for the different schemes. In addition, the sum-rates without any optimization are also shown for all the schemes, meaning that the maximum available transmit powers are used, and $\eta = 0.5$. This illustrates the gains achieved by optimizing the transmit powers.


\begin{table}[!t]
\renewcommand{\arraystretch}{1.3}
\caption{Example parameters for the considered system.}
\label{table:def_param}
\centering
\begin{tabular}{|c||c|}
\hline
\textbf{Parameter} & \textbf{Value}\\
\hline
Total number of TX antennas at the AN ($N_t$) & 200\\
\hline
Total number of RX antennas at the AN ($N_r$) & 100\\
\hline
Number of transmitted backhaul signal streams ($M^\mathit{BH}_\mathit{t}$)& 6\\
\hline
Number of received backhaul signal streams ($M^\mathit{BH}_\mathit{r}$)& 12\\
\hline
Number of DL UEs ($D$) & 10\\
\hline
Number of UL UEs ($U$) & 10\\
\hline
Noise floor in all the receivers ($\sigma_n^2$)& $-90$ dBm\\
\hline
Path loss between the AN and the UEs ($L_\mathit{UE}$) & 80 dB\\
\hline
Path loss between UL and DL UEs ($L_\mathit{UD}$) & 70 dB\\
\hline
Path loss between the AN and the BN ($L_\mathit{BH}$) & 80 dB\\
\hline
Maximum transmit power of the AN ($P_\mathit{AN}$) & +30 dBm\\
\hline
Maximum transmit power of an UE ($P_\mathit{UE}$) & +25 dBm\\
\hline
Maximum transmit power of the BN ($P^\mathit{BH}_\mathit{d,max}$) & +40 dBm\\
\hline
The amount of SI cancellation in the AN & 120 dB\\
\hline
Ratio between UL and DL data rates ($\rho_\mathit{min}$--$\rho_\mathit{max}$) & 15--30 \% \\
\hline
\end{tabular}
\figlowerspace
\end{table}

\begin{figure*}
        \centering
        \begin{subfigure}[t]{0.49\textwidth}
                \includegraphics[width=\textwidth]{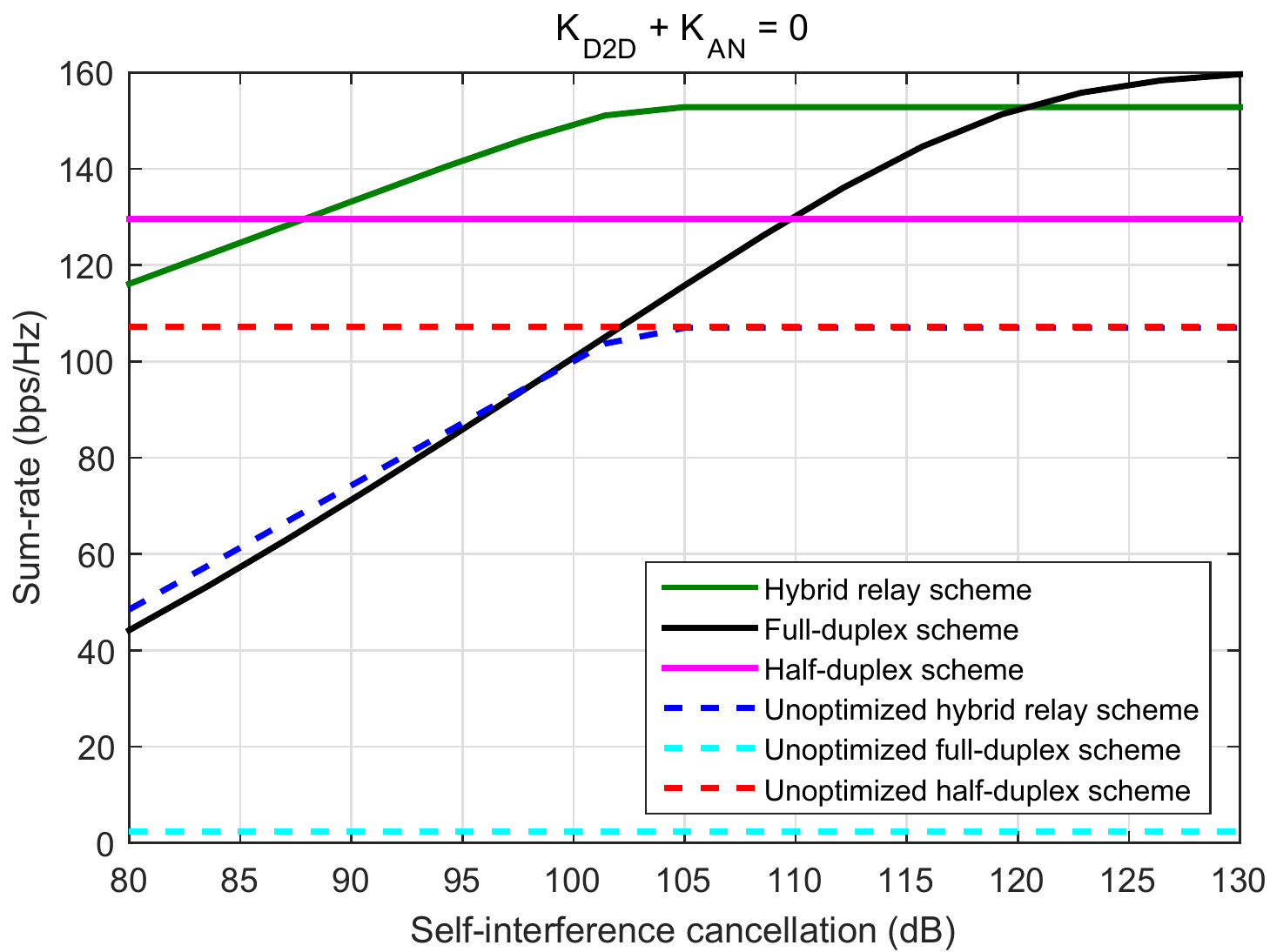}	
								\caption{}
                \label{fig:sic_varied_k_0}
        \end{subfigure}%
        ~
        \begin{subfigure}[t]{0.49\textwidth}
                \includegraphics[width=\textwidth]{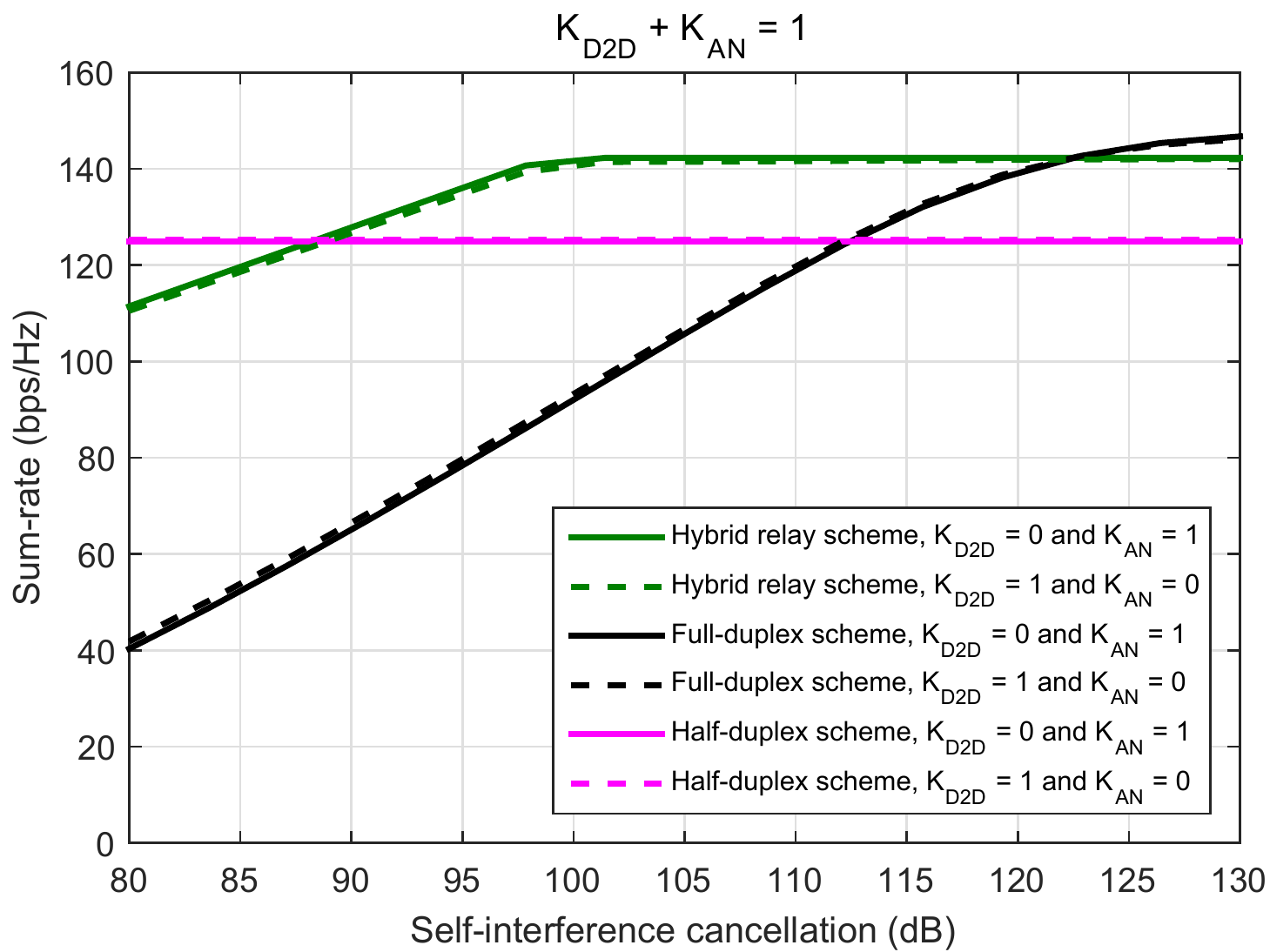}
								\caption{}
                \label{fig:sic_varied_k_1}
        \end{subfigure}
        \caption{The maximum sum-rates for the different schemes with respect to the amount of self-interference cancellation with (a) no intra-cell traffic, (b) one UE pair communicating within the cell.}
\label{fig:sic}
\figlowerspace
\end{figure*}

Figure~\ref{fig:sic}(\subref{fig:sic_varied_k_0}) shows the maximum sum-rates for different amounts of SI cancellation, with the other parameters being set according to Table~\ref{table:def_param}. In this case, it is assumed that none of the UEs is communicating within the cell, i.e., $K_\mathit{D2D} + K_\mathit{AN} = 0$. The sum-rate of the half-duplex scheme is obviously constant, as it does not depend on the SI cancellation ability of the AN. With these parameters, the optimized half-duplex scheme provides the highest sum-rate when the amount of SI cancellation is less than 88~dB. This is the threshold value for the SI cancellation performance, under which full-duplex communications is not beneficial in any form.

When the SI cancellation performance is between 88~dB and 120~dB, the hybrid relay scheme is the optimal solution. Within this range, the full-duplex capability of the AN is beneficial, while it is still not enough to compensate for the negative effect of IUI. Thus, the hybrid relay scheme, which utilizes the full-duplex capability of the AN while avoiding the IUI by design, provides the highest sum-rate. However, when the amount of SI cancellation rises above 120~dB, the pure full-duplex scheme provides the best performance. This is the point, after which the optimal solution is to do as much of the transmissions in full-duplex mode as possible. Thus, even at the cost of significant IUI, the AN should serve the UL and DL users at the same time to maximize the sum-rate.

It is also evident that using the maximum transmit powers provides clearly an inferior sum-rate. The full-duplex scheme obtains the worst performance without optimization, since it can hardly provide any data rate then. This is due to the fact that in this case the AN transmit power is simply divided evenly between the backhaul and DL transmissions, which results in an extremely low DL data rate. The hybrid relay and half-duplex schemes obtain reasonable sum-rates also without optimization, but they are still heavily outperformed by their optimized counterparts.

Figure~\ref{fig:sic}(\subref{fig:sic_varied_k_1}) shows then the sum-rates for the optimized schemes when $K_\mathit{D2D} + K_\mathit{AN} = 1$, i.e., when there is one UE pair communicating only within the cell. It can be observed that the manner in which the intra-cell traffic is handled does not significantly affect the sum-rate. However, overall, the sum-rates for this situation are slightly lower than when $K_\mathit{D2D} + K_\mathit{AN} = 0$. For the case with $ K_\mathit{AN} = 1$, this is caused by the fact that the intra-cell traffic is counted only once when calculating the sum-rate. In the case of Fig.~\ref{fig:sic}(\subref{fig:sic_varied_k_0}), the data rate for both of these UEs is included in the sum-rate, resulting in a higher result. However, if the backhaul capacity of the AN is low, then having part of the UEs communicating only within the cell will result in a higher sum-rate, as will be shown later.

When $K_\mathit{D2D} = 1$, the lower sum-rate is explained by the generally lower SINR of the D2D link, caused by the limited UE transmit power. Thus, even though the AN has more power available to serve the remaining DL UEs, the low rate of the D2D link is enough to reverse the positive effects. All in all, for these fully optimized schemes, there is essentially no difference in the relative performances between the cases of $K_\mathit{D2D} = 1$ and $K_\mathit{AN} = 1$. Hence, with these system parameter values, it does not matter whether the intra-cell traffic is handled via the AN or directly D2D.

\begin{figure*}
        \centering
        \begin{subfigure}[t]{0.49\textwidth}
                \includegraphics[width=\textwidth]{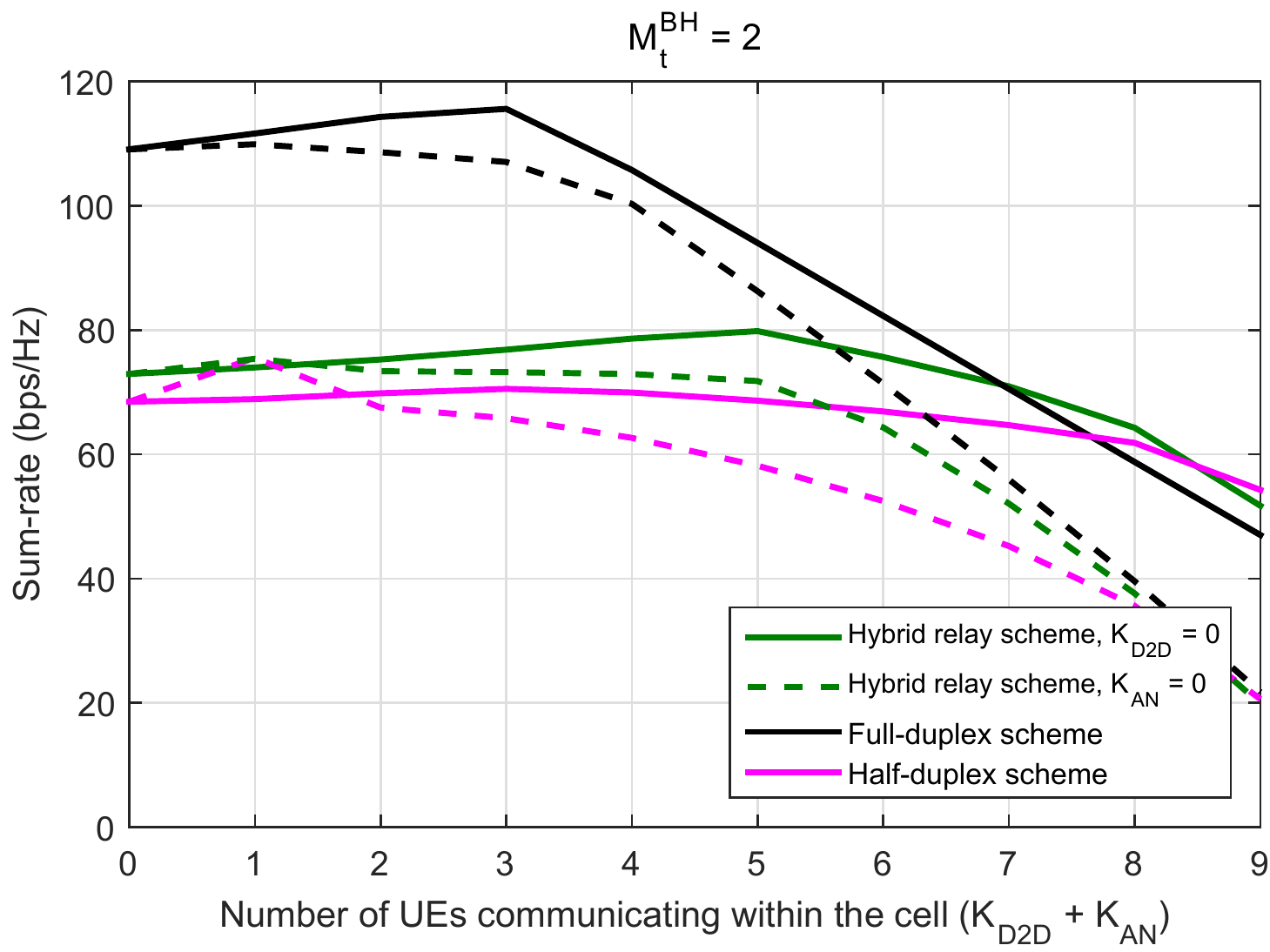}	
								\caption{}
                \label{fig:K_varied_N_BH_2}
        \end{subfigure}%
        ~
        \begin{subfigure}[t]{0.49\textwidth}
                \includegraphics[width=\textwidth]{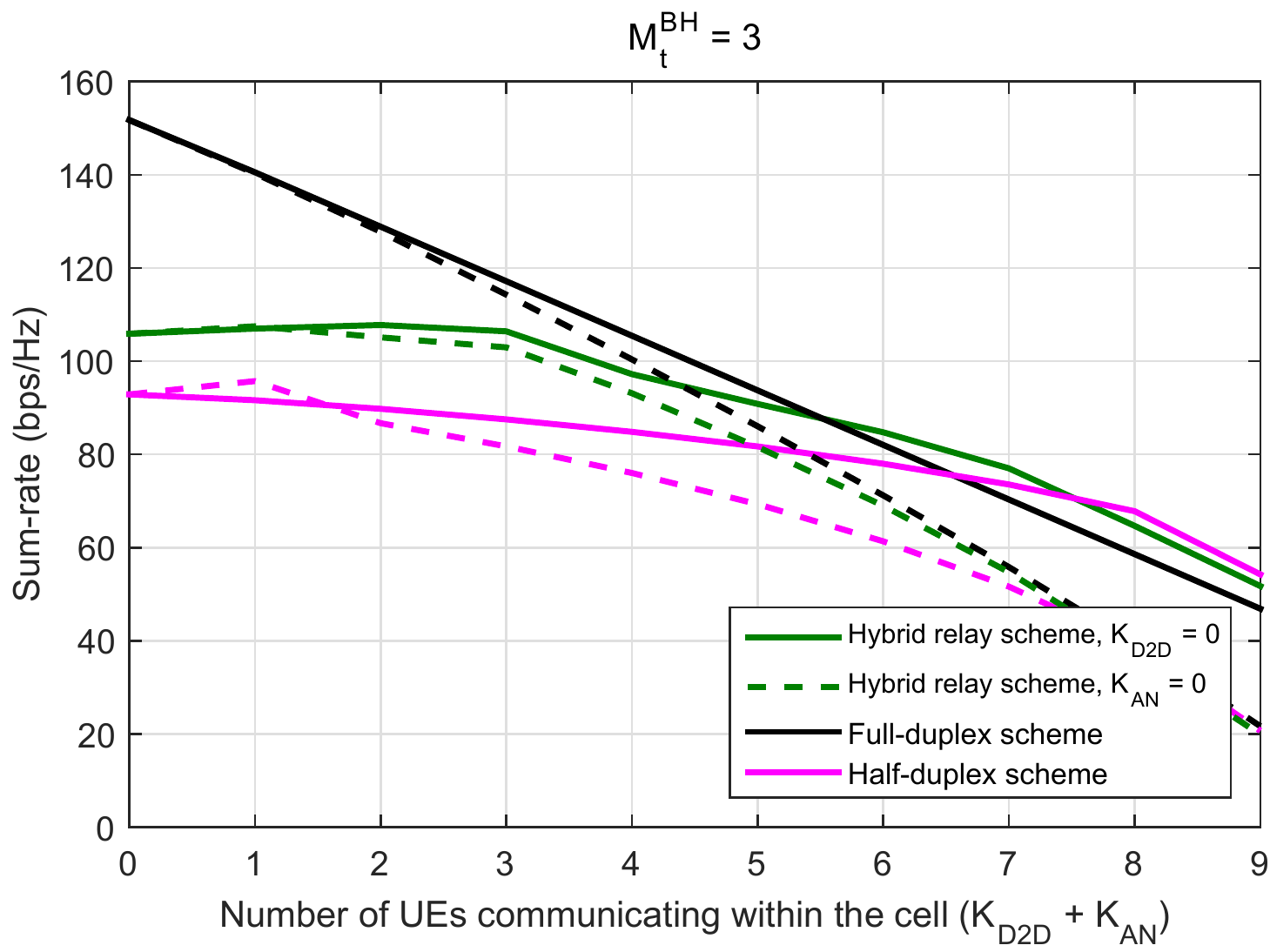}
								\caption{}
                \label{fig:K_varied_N_BH_3}
        \end{subfigure}
        \caption{The maximum sum-rates for the different schemes with respect to the number of UE pairs communicating within the cell, when the number of transmitted backhaul signal streams is (a) 2 and (b) 3.}
\label{fig:K}
\end{figure*}

Another significant aspect for the performance of the considered AN is the capacity of the backhaul link, which is greatly dependent on the number of signal streams. For this reason, Figs.~\ref{fig:K}(\subref{fig:K_varied_N_BH_2}) and~\ref{fig:K}(\subref{fig:K_varied_N_BH_3}) show the achievable sum-rates when the number of transmitted backhaul signal streams is set to 2 and 3, respectively. The sum-rates are plotted with respect to the number of UE pairs communicating within the considered cell. The number of received backhaul signal streams is always set to be twice the number of transmitted signal streams to allow for a higher data rate in the DL.

When there are only two transmitted signal streams, the backhaul link is heavily bottlenecking the achievable sum-rate of the cell. For this reason, the intuitive assessment is that the more of the UEs are communicating within the cell, the higher is the sum-rate that can be achieved. Without D2D transmissions ($K_\mathit{D2D} = 0$), this seems to be the case to certain extent. For the full-duplex scheme, the highest sum-rate is achieved when there are 3 UE pairs communicating within the cell (i.e., $K_\mathit{AN} = 3$). Beyond this point, the sum-rate starts to decrease since the backhaul connection can already support the traffic requirements of the rest of the UEs. This phenomenon occurs also with the hybrid relay and half-duplex schemes, with the former obtaining the highest rate when $K_\mathit{AN} = 5$, and the latter when $K_\mathit{AN} = 3$. The reason for the different optimal point of the hybrid relay scheme is in the way the capacity is divided between the UL, DL, and backhaul link. Namely, each of the three schemes have different characteristics for the achievable rates of the different links, which in turn define the optimal ratio between intra-cell and outgoing traffic.

The sum-rate behaves somewhat differently when the intra-cell traffic is handled by D2D transmissions ($K_\mathit{AN} = 0$). When observing Fig.~\ref{fig:K}(\subref{fig:K_varied_N_BH_2}), it can be seen that in this case the sum-rate of all the schemes is maximized when $K_\mathit{D2D} = 1$. This is explained by the relationship between the additional interference produced by the D2D transmission and the amount of resources it frees in the AN. When there is only one UE pair communicating in a D2D fashion, the additional IUI only affects the DL UEs, while with more than one D2D UE pair the different D2D transmissions interfere also with each other. For this reason, under a low-rate backhaul connection, the most optimal value for $K_\mathit{D2D}$ is 1. Moreover, with the hybrid relay and half-duplex schemes, in such a case it is in fact better to handle the intra-cell traffic with D2D communication instead of routing it via the AN.

The same general conclusions apply also to the case with three transmitted backhaul signal streams, as can be observed in Fig.~\ref{fig:K}(\subref{fig:K_varied_N_BH_3}), with the difference that now the optimal amounts of intra-cell UE pairs for the different schemes are obviously smaller, due to the higher capacity of the backhaul link. Overall, it should also be noted that, for both $M_\mathit{t}^\mathit{BH} = 2$ and $M_\mathit{t}^\mathit{BH} = 3$, the full-duplex scheme provides the highest sum-rate of all the schemes, as long as there is only a reasonable amount of intra-cell communication. This indicates that the full-duplex scheme is the best solution with scarce backhauling resources, since it can utilize a small number of backhaul signal streams more efficiently than the other solutions.

\begin{figure*}
        \centering
        \begin{subfigure}[t]{0.49\textwidth}
                \includegraphics[width=\textwidth]{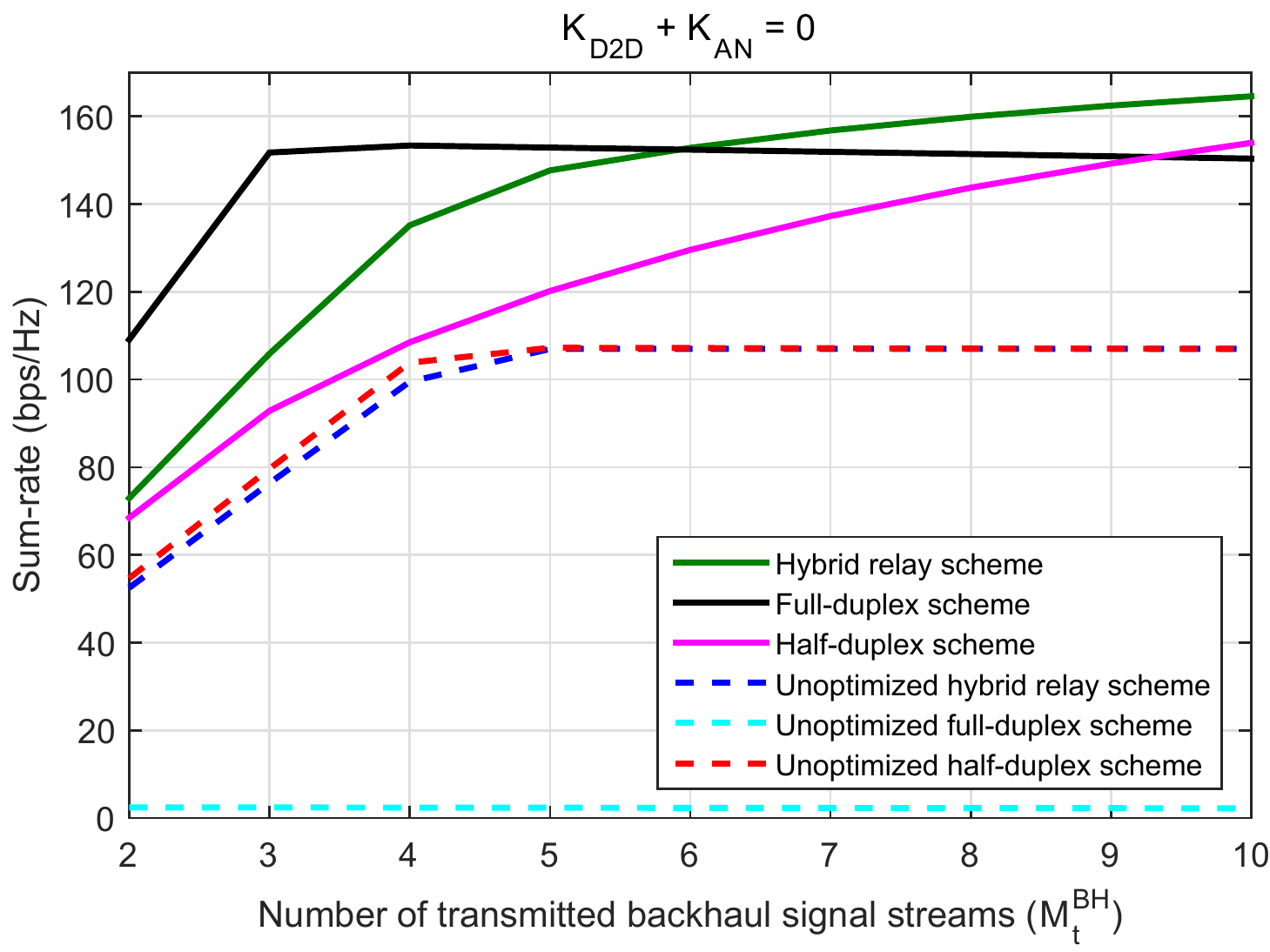}	
								\caption{}
                \label{fig:N_BH_varied_K_0}
        \end{subfigure}%
        ~
        \begin{subfigure}[t]{0.49\textwidth}
                \includegraphics[width=\textwidth]{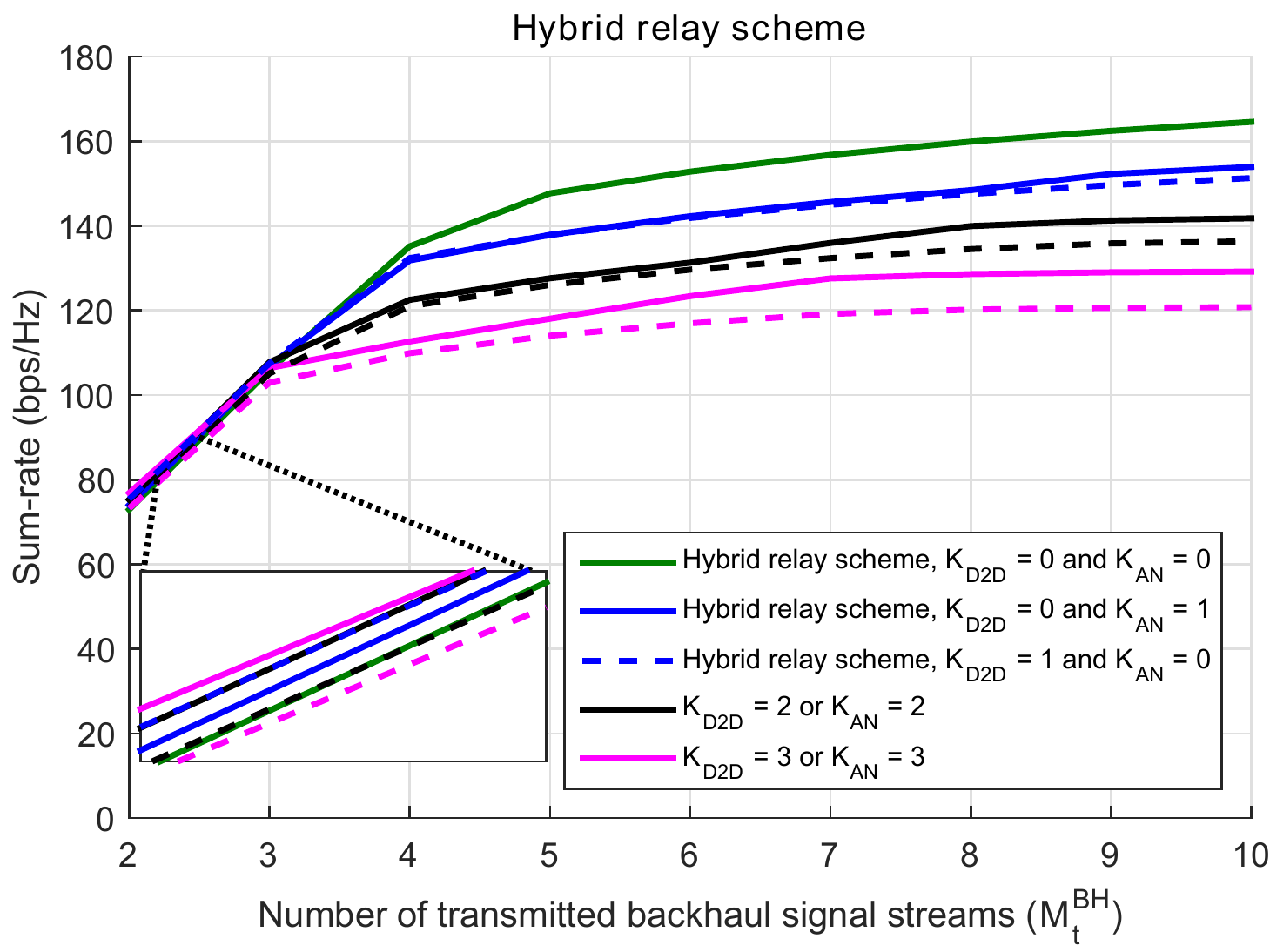}
								\caption{}
                \label{fig:N_BH_varied_RL}
        \end{subfigure}
        \caption{The maximum sum-rates with respect to the number of transmitted backhaul signal streams (a) for the different schemes with no intra-cell traffic, (b) for the hybrid relay scheme with various amounts of UE pairs communicating within the cell.}
\label{fig:N_BH}
\figlowerspace
\end{figure*}

To investigate the effect of the backhaul link capacity further, Fig.~\ref{fig:N_BH}(\subref{fig:N_BH_varied_K_0}) shows the sum-rates of the different schemes with respect to the number of transmitted backhaul signal streams. In this case, it is assumed that there is no intra-cell traffic, i.e., $K_\mathit{D2D}+K_\mathit{AN} = 0$. As can be expected, if the number of signal streams is very small, the capacity of the backhaul link is bottlenecking the sum-rate of the cell, which is also clearly visible in the figure. There are also differences in how the capacity of the backhaul link affects the different schemes. Similar to the earlier observations, with a low-rate backhaul link, the full-duplex scheme clearly outperforms all the other schemes, since it can utilize the backhaul resources more efficiently. However, its performance saturates already with 3 or more transmitted backhaul signal streams, and thus it is not the best solution when the data rate of the backhaul link is high.

The half-duplex and hybrid relay schemes, on the other hand, benefit greatly from any increase in the backhaul link capacity. They perform rather poorly when the backhaul capacity is low but can obtain a high sum-rate when given more backhauling resources. Especially, the hybrid relay scheme is the best solution when there are 6 or more transmitted backhaul signal streams, while also the half-duplex scheme outperforms the full-duplex scheme with 10 or more signal streams.

The sum-rates of the unoptimized schemes are again lower than those of the optimized schemes, as can be expected. The effect of optimization on the hybrid relay and half-duplex schemes is not as significant when the backhaul capacity is the bottleneck. However, the sum-rate of the unoptimized schemes saturates with 5 or more backhaul signal streams, while the optimized schemes can improve their sum-rate even after this point. Hence, with a high-rate backhaul link, the significance of optimizing the transmit powers is especially emphasized.


To obtain further insight into how the amount of intra-cell traffic affects the sum-rate with different backhauling capacities, Fig.~\ref{fig:N_BH}(\subref{fig:N_BH_varied_RL}) shows the sum-rate of the hybrid relay scheme with respect to the number of transmitted backhaul signal streams when $K_\mathit{D2D}+K_\mathit{AN} = \{0\text{, }1\text{, }2\text{, }3 \}$. It can be observed that, when the capacity of the backhaul link is the bottleneck, having a larger amount of intra-cell traffic provides a higher sum-rate, as long as the traffic is routed via the AN. This is intuitively clear, since in this case every UE pair communicating within the cell will provide a certain data rate without requiring any backhauling. Furthermore, with scarce backhauling resources and only one intra-cell UE pair, it can be observed that it is better to handle the intra-cell traffic with D2D transmissions, since the case with $K_\mathit{D2D} = 1 $ provides a higher sum-rate than the one with $K_\mathit{AN} = 1$. This observation is in line with the deductions made from Figs.~\ref{fig:K}(\subref{fig:K_varied_N_BH_2}) and~\ref{fig:K}(\subref{fig:K_varied_N_BH_3}).

However, when the number of backhaul signal streams is 4 or higher, a larger amount of intra-cell traffic will in fact decrease the sum-rate. The reason for this is that all the intra-cell data is generated and consumed within the cell, and thereby it only has a onefold contribution to the sum-rate, as discussed earlier. On the other hand, if the UE pairs were communicating with the outside world, the respective data rates of both the transmitting and receiving UEs would be counted towards the sum-rate, resulting in a much larger contribution. Thus, with a high-capacity backhaul link, a large amount of intra-cell traffic is not beneficial when using this particular metric.

Overall, the obtained results indicate that utilizing the full-duplex capability of the AN typically results in a higher sum-rate than performing all the communication in half-duplex mode. In particular, the results indicate that as long as the total amount of SI cancellation is at least 90~dB, either the hybrid relay or the full-duplex scheme provides the highest sum-rate. This level of SI cancellation can already be reached with actual demonstrator implementations, such as those reported in \cite{Korpi15d,Heino15a}. However, it is important to note that the transmit powers of the schemes must be optimized in order to reach the aforementioned performance levels. Simply using the maximum transmit powers does not result in any performance gain when utilizing the full-duplex capability of the AN. An additional benefit of optimizing the transmit powers is the lower amount of radiated power, which reduces the energy consumption of the network while also improving the spectral efficiency.

\section{Conclusion}
\label{sec:conc}

This article studied and analyzed a radio access system where a full-duplex access node is wirelessly backhauling itself while serving legacy half-duplex UEs. In order to maximize the spectral efficiency, all of the transmissions were assumed to be done using the same center-frequency, meaning that also the backhauling is done in full-duplex mode. The proposed system was then evaluated in terms of the sum-rate that could be achieved under the constraint that all the data was backhauled wirelessly. The achievable sum-rates of the full-duplex access node were then compared to corresponding half-duplex and hybrid relay access nodes. After obtaining the sum-rate values by numerical optimization using typical system parameter values, it was observed that under most circumstances the highest sum-rate is provided by the hybrid relay scheme, which essentially means that the access node acts as a full-duplex relay between the UEs and the backhaul node. In addition, the results showed that the half-duplex scheme is under most circumstances the worst option in terms of the achievable sum-rate. Hence, in order to obtain the highest possible sum-rate, the full-duplex capability of the access node should be utilized, leading to improved spectral efficiency and reduced energy consumption of the network when combined with proper transmit power optimization.

\appendix

\subsection{Derivation of the Normalization Factor for the Zero-Forcing Precoding Matrix}
\label{app:zf_norm}

In order to maintain the power of each symbol during the precoding procedure, the column wise norms of $\mathbf{W}$ corresponding to the actual data symbols must be normalized to 1. This ensures that, when considering the precoded transmit data, the total power corresponding to each individual data symbol is not affected by precoding. Thus, we get the following constraint:
\begin{align}
	\operatorname{E}\left[\norm{\mathbf{w}_k}^2\right] = 1 \label{eq:W_constr}\text{,}
\end{align}
where $\mathbf{w}_k$ is the $k$th column of $\mathbf{W}$. The unnormalized precoding matrix is defined as
\begin{align}
	\widetilde{\mathbf{W}} = \mathbf{H}^H\left(\mathbf{H}\mathbf{H}^H\right)^{-1} \text{.} \label{eq:w_unnorm}
\end{align}
Let us now investigate its $k$th column, denoted by $\widetilde{\mathbf{w}}_k$. Using basic matrix algebra, we can express the expected squared norm as
\begin{align}
	\operatorname{E}\left[\norm{\widetilde{\mathbf{w}}_k}^2\right] =\operatorname{E}\left[\norm{\widetilde{\mathbf{W}}\mathbf{e}_k}^2\right] = \operatorname{E}\left[\left(\widetilde{\mathbf{W}}\mathbf{e}_k\right)^H\left(\widetilde{\mathbf{W}}\mathbf{e}_k\right)\right] = \mathbf{e}_k^H \operatorname{E}\left[\widetilde{\mathbf{W}}^H\widetilde{\mathbf{W}}\right]\mathbf{e}_k\text{,} \label{eq:norm_eq_mod}
\end{align}
where $\mathbf{e}_k$ is the standard unit vector having 1 in the $k$th element and zeros elsewhere. Substituting \eqref{eq:w_unnorm} into \eqref{eq:norm_eq_mod}, we get
\begin{align}
	\mathbf{e}_k^H \operatorname{E}\left[\widetilde{\mathbf{W}}^H\widetilde{\mathbf{W}}\right]\mathbf{e}_k = \mathbf{e}_k^H \operatorname{E}\left[\left(\mathbf{H}\mathbf{H}^H\right)^{-1} \mathbf{H} \mathbf{H}^H \left(\mathbf{H}\mathbf{H}^H\right)^{-1}\right]\mathbf{e}_k = \operatorname{E}\left[\left\{\left(\mathbf{H}\mathbf{H}^H\right)^{-1}\right\}_\mathit{kk}\right] \text{.}
\end{align}

Now, in order to simplify the above expression, we assume that also the effective SI channel experiences Rayleigh fading, which is a relatively accurate assumption when there is a certain level of SI cancellation before the total received signal is decoded \cite{Duarte12}. Thus, we can write $\mathbf{H}_s \sim \mathcal{CN}(0,\mathbf{L}_\mathit{SI})$, and as a result also $\mathbf{H} \sim \mathcal{CN}(0,\mathbf{L}_\mathit{tot})$, where $\mathbf{L}_\mathit{tot} = \operatorname{diag}\left(\operatorname{diag}\left(\mathbf{L}\right)\text{, }\operatorname{diag}\left(\mathbf{L}_\mathit{SI}\right)\right)$. This allows us to write $\mathbf{H} = \mathbf{L}_\mathit{tot}^{1/2} \breve{\mathbf{H}}$, where $\breve{\mathbf{H}}\sim \mathcal{CN}(0,\mathbf{I})$. Then, we get
\begin{align}
	\operatorname{E}\left[\left\{\left(\mathbf{H}\mathbf{H}^H\right)^{-1}\right\}_\mathit{kk}\right] &= \operatorname{E}\left[\left\{\left(\mathbf{L}_\mathit{tot}^{1/2} \breve{\mathbf{H}}\left(\mathbf{L}_\mathit{tot}^{1/2} \breve{\mathbf{H}}\right)^H\right)^{-1}\right\}_\mathit{kk}\right] = \operatorname{E}\left[\left\{\mathbf{L}_\mathit{tot}^{-1/2}\left( \breve{\mathbf{H}}\breve{\mathbf{H}}^H\right)^{-1} \mathbf{L}_\mathit{tot}^{-1/2}\right\}_\mathit{kk}\right]\nonumber\\
	&= \frac{1}{L_k}\operatorname{E}\left[\left\{\left( \breve{\mathbf{H}}\breve{\mathbf{H}}^H\right)^{-1}\right\}_\mathit{kk}\right]\text{,} \label{eq:norm_ch}
\end{align}
where $L_k$ is the $k$th diagonal of $\mathbf{L}_\mathit{tot}$. Furthermore, when $N_t$ is large, the $k$th diagonal of $\left(\breve{\mathbf{H}}\breve{\mathbf{H}}^H\right)^{-1}$ can be approximated by the average value of all of its diagonals. Thus, we can rewrite \eqref{eq:norm_ch} as follows:
\begin{align}
	\frac{1}{L_k}\operatorname{E}\left[\left\{\left(\breve{\mathbf{H}}\breve{\mathbf{H}}^H\right)^{-1}\right\}_\mathit{kk}\right] \approx \frac{1}{\left(M_t+N_r\right)L_\mathit{k}}\operatorname{E}\left[\operatorname{Trace}\left\{\left(\breve{\mathbf{H}}\breve{\mathbf{H}}^H\right)^{-1}\right\}\right]
\end{align}
Finally, since $\breve{\mathbf{H}} \sim \mathcal{CN}(0,\mathbf{I})$, $\breve{\mathbf{H}}\breve{\mathbf{H}}^H$ is an $\left(M_t+N_r\right)\times \left(M_t+N_r\right)$ central Wishart matrix with $N_t$ degrees of freedom, and the following property holds \cite[Lemma 2.10]{Tulino04}:
\begin{align}
	\operatorname{E}\left[\operatorname{Trace}\left\{\left(\breve{\mathbf{H}}\breve{\mathbf{H}}^H\right)^{-1}\right\}\right] = \frac{M_t+N_r}{N_t-M_t-N_r}\text{.} \label{eq:wishart}
\end{align}
Thus, we have
\begin{align}
	\operatorname{E}\left[\norm{\widetilde{\mathbf{w}}_k}^2\right] = \frac{1}{L_\mathit{k}\left(N_t-M_t-N_r\right)}\text{.}
\end{align}
Hence, it is clear that if
\begin{align}
	\mathbf{w}_k = \sqrt{L_\mathit{k}\left(N_t-M_t-N_r\right)}\widetilde{\mathbf{w}}_k \text{,} \label{eq:w_k_norm}
\end{align}
then
\begin{align}
	\operatorname{E}\left[\norm{\mathbf{w}_k }^2\right] = \operatorname{E}\left[\norm{\sqrt{L_\mathit{k}\left(N_t-M_t-N_r\right)}\widetilde{\mathbf{w}}_k }^2\right] = L_\mathit{k}\left(N_t-M_t-N_r\right)\operatorname{E}\left[\norm{\widetilde{\mathbf{w}}_k }^2\right] = 1 \label{eq:w_norm_proof}\text{,}
\end{align}
and the constraint given in \eqref{eq:W_constr} is fulfilled. Thereby, this gives us the normalization factors as follows:
\begin{align}
	\lambda_k = \sqrt{L_\mathit{k}\left(N_t-M_t-N_r\right)} \text{.} \label{eq:c_norm}
\end{align}
In order to multiply the $k$th column of $\widetilde{\mathbf{W}}$ by $\lambda_k$ and transform it into the required dimensions, the normalization factors for $k = \left\{1\text{, }2\text{,}\ldots\text{, }M_t\right\}$ are used to form the $\left(M_t+N_r\right) \times M_t$ diagonal matrix $\mathbf{\Lambda}$, which is used as shown in \eqref{eq:precoder}.

In half-duplex mode, $\mathbf{H} = \mathbf{H}_t$, which results in normalization factors of the form $\lambda_k = \sqrt{L_\mathit{k}\left(N_t-M_t\right)}$ after an identical derivation. In a similar manner, the normalization factors for the receiver precoder in all the communication modes are $\lambda_k = \sqrt{L_\mathit{k}\left(N_r-M_r\right)}$, where $M_r$ is the number of parties transmitting to the AN.

\bibliographystyle{IEEEtran}
\bibliography{IEEEref}

\begin{thebibliography}{10}
\providecommand{\url}[1]{#1}
\csname url@samestyle\endcsname
\providecommand{\newblock}{\relax}
\providecommand{\bibinfo}[2]{#2}
\providecommand{\BIBentrySTDinterwordspacing}{\spaceskip=0pt\relax}
\providecommand{\BIBentryALTinterwordstretchfactor}{4}
\providecommand{\BIBentryALTinterwordspacing}{\spaceskip=\fontdimen2\font plus
\BIBentryALTinterwordstretchfactor\fontdimen3\font minus
  \fontdimen4\font\relax}
\providecommand{\BIBforeignlanguage}[2]{{%
\expandafter\ifx\csname l@#1\endcsname\relax
\typeout{** WARNING: IEEEtran.bst: No hyphenation pattern has been}%
\typeout{** loaded for the language `#1'. Using the pattern for}%
\typeout{** the default language instead.}%
\else
\language=\csname l@#1\endcsname
\fi
#2}}
\providecommand{\BIBdecl}{\relax}
\BIBdecl

\bibitem{Duarte12}
M.~Duarte, C.~Dick, and A.~Sabharwal, ``Experiment-driven characterization of
  full-duplex wireless systems,'' \emph{IEEE Transactions on Wireless
  Communications}, vol.~11, no.~12, pp. 4296--4307, Dec. 2012.

\bibitem{Bliss07}
D.~Bliss, P.~Parker, and A.~Margetts, ``Simultaneous transmission and reception
  for improved wireless network performance,'' in \emph{Proc. IEEE/SP 14th
  Workshop on Statistical Signal Processing}, Aug. 2007, pp. 478--482.

\bibitem{Day12}
B.~Day, A.~Margetts, D.~Bliss, and P.~Schniter, ``Full-duplex bidirectional
  {MIMO}: Achievable rates under limited dynamic range,'' \emph{IEEE
  Transactions on Signal Processing}, vol.~60, no.~7, pp. 3702--3713, Jul.
  2012.

\bibitem{Choi10}
J.~I. Choi, M.~Jain, K.~Srinivasan, P.~Levis, and S.~Katti, ``Achieving single
  channel full duplex wireless communication,'' in \emph{Proc. 16th Annual
  International Conference on Mobile Computing and Networking}, Sep. 2010, pp.
  1--12.

\bibitem{Jain11}
M.~Jain, J.~I. Choi, T.~Kim, D.~Bharadia, S.~Seth, K.~Srinivasan, P.~Levis,
  S.~Katti, and P.~Sinha, ``Practical, real-time, full duplex wireless,'' in
  \emph{Proc. 17th Annual International Conference on Mobile computing and
  Networking}, Sep. 2011, pp. 301--312.

\bibitem{Korpi14c}
D.~Korpi, T.~Riihonen, V.~Syrj\"{a}l\"{a}, L.~Anttila, M.~Valkama, and
  R.~Wichman, ``Full-duplex transceiver system calculations: analysis of {ADC}
  and linearity challenges,'' \emph{IEEE Transactions on Wireless
  Communications}, vol.~13, no.~7, pp. 3821--3836, Jul. 2014.

\bibitem{Goyal15a}
S.~Goyal, P.~Liu, S.~S. Panwar, R.~A. Difazio, R.~Yang, and E.~Bala, ``Full
  duplex cellular systems: will doubling interference prevent doubling
  capacity?'' \emph{IEEE Communications Magazine}, vol.~53, no.~5, pp.
  121--127, May 2015.

\bibitem{Heino15a}
M.~Heino, D.~Korpi, T.~Huusari, E.~Antonio-Rodr\'{i}guez,
  S.~Venkatasubramanian, T.~Riihonen, L.~Anttila, C.~Icheln, K.~Haneda,
  R.~Wichman, and M.~Valkama, ``Recent advances in antenna design and
  interference cancellation algorithms for in-band full-duplex relays,''
  \emph{IEEE Communications Magazine}, vol.~53, no.~5, pp. 91--101, May 2015.

\bibitem{Bharadia13}
D.~Bharadia, E.~McMilin, and S.~Katti, ``Full duplex radios,'' in \emph{Proc.
  SIGCOMM'13}, Aug. 2013, pp. 375--386.

\bibitem{Korpi15d}
D.~Korpi, Y.-S. Choi, T.~Huusari, S.~Anttila, L.~Talwar, and M.~Valkama,
  ``Adaptive nonlinear digital self-interference cancellation for mobile inband
  full-duplex radio: algorithms and {RF} measurements,'' in \emph{Proc. IEEE
  Global Communications Conference (GLOBECOM)}, Dec. 2015.

\bibitem{Day122}
B.~Day, A.~Margetts, D.~Bliss, and P.~Schniter, ``Full-duplex {MIMO} relaying:
  Achievable rates under limited dynamic range,'' \emph{IEEE Journal on
  Selected Areas in Communications}, vol.~30, no.~8, pp. 1541--1553, Sep. 2012.

\bibitem{Riihonen11}
T.~Riihonen, S.~Werner, and R.~Wichman, ``Mitigation of loopback
  self-interference in full-duplex {MIMO} relays,'' \emph{IEEE Transactions on
  Signal Processing}, vol.~59, no.~12, pp. 5983--5993, Dec. 2011.

\bibitem{Riihonen13}
T.~Riihonen, M.~Vehkapera, and R.~Wichman, ``Large-system analysis of rate
  regions in bidirectional full-duplex {MIMO} link: Suppression versus
  cancellation,'' in \emph{Proc. 47th Annual Conference on Information Sciences
  and Systems (CISS)}, Mar. 2013, pp. 1--6.

\bibitem{Korpi15a}
D.~Korpi, T.~Riihonen, and M.~Valkama, ``Achievable rate regions and
  self-interference channel estimation in hybrid full-duplex/half-duplex radio
  links,'' in \emph{Proc. 49th Annual Conference on Information Sciences and
  Systems (CISS)}, Mar. 2015, pp. 1--6.

\bibitem{Korpi15c}
D.~Korpi, T.~Riihonen, K.~Haneda, K.~Yamamoto, and M.~Valkama, ``Achievable
  transmission rates and self-interference channel estimation in hybrid
  full-duplex/half-duplex {MIMO} relaying,'' in \emph{Proc. IEEE 82nd Vehicular
  Technology Conference (VTC Fall)}, Sep. 2015.

\bibitem{Aggarwal12}
V.~Aggarwal, M.~Duarte, A.~Sabharwal, and N.~Shankaranarayanan, ``Full- or
  half-duplex? {A} capacity analysis with bounded radio resources,'' in
  \emph{Proc. Information Theory Workshop}, Sep. 2012, pp. 207--211.

\bibitem{Anttila13}
L.~Anttila, D.~Korpi, V.~Syrj\"{a}l\"{a}, and M.~Valkama, ``Cancellation of
  power amplifier induced nonlinear self-interference in full-duplex
  transceivers,'' in \emph{Proc. 47th Asilomar Conference on Signals, Systems
  and Computers}, Nov. 2013, pp. 1193--1198.

\bibitem{Korpi14d}
D.~Korpi, L.~Anttila, V.~Syrj\"{a}l\"{a}, and M.~Valkama, ``Widely linear
  digital self-interference cancellation in direct-conversion full-duplex
  transceiver,'' \emph{IEEE Journal on Selected Areas in Communications},
  vol.~32, no.~9, pp. 1674--1687, Sep. 2014.

\bibitem{Ahmed13}
E.~Ahmed, A.~M. Eltawil, and A.~Sabharwal, ``Self-interference cancellation
  with nonlinear distortion suppression for full-duplex systems,'' in
  \emph{Proc. 47th Asilomar Conference on Signals, Systems and Computers}, Nov.
  2013, pp. 1199--1203.

\bibitem{Kaufman13}
B.~Kaufman, J.~Lilleberg, and B.~Aazhang, ``An analog baseband approach for
  designing full-duplex radios,'' in \emph{Proc. 47th Asilomar Conference on
  Signals, Systems and Computers}, Nov. 2013, pp. 987--991.

\bibitem{Liu14}
Y.~Liu, X.~Quan, W.~Pan, S.~Shao, and Y.~Tang, ``Nonlinear distortion
  suppression for active analog self-interference cancellers in full duplex
  wireless communication,'' in \emph{Proc. Globecom Workshops}, Dec. 2014, pp.
  948--953.

\bibitem{Huusari15}
T.~Huusari, Y.-S. Choi, P.~Liikkanen, D.~Korpi, S.~Talwar, and M.~Valkama,
  ``Wideband self-adaptive {RF} cancellation circuit for full-duplex radio:
  Operating principle and measurements,'' in \emph{Proc. IEEE 81st Vehicular
  Technology Conference (VTC Spring)}, May 2015.

\bibitem{Sabharwal14}
A.~Sabharwal, P.~Schniter, D.~Guo, D.~Bliss, S.~Rangarajan, and R.~Wichman,
  ``In-band full-duplex wireless: Challenges and opportunities,'' \emph{IEEE
  Journal on Selected Areas in Communications}, vol.~32, no.~9, Oct. 2014.

\bibitem{Everett11}
E.~Everett, M.~Duarte, C.~Dick, and A.~Sabharwal, ``Empowering full-duplex
  wireless communication by exploiting directional diversity,'' in \emph{Proc.
  45th Asilomar Conference on Signals, Systems and Computers}, Nov. 2011, pp.
  2002--2006.

\bibitem{nsn13}
{Nokia Solutions and Networks}, ``{TD-LTE} frame configuration primer,'' Nov.
  2013, white paper.

\bibitem{Tabassum15a}
H.~Tabassum, A.~H. Sakr, and E.~Hossain, ``Massive {MIMO}-enabled wireless
  backhauls for full-duplex small cells,'' in \emph{Proc. IEEE Global
  Communications Conference (GLOBECOM)}, Dec. 2015, pp. 1--6.

\bibitem{Pitaval15a}
R.-A. Pitaval, O.~Tirkkonen, R.~Wichman, K.~Pajukoski, E.~Lahetkangas, and
  E.~Tiirola, ``Full-duplex self-backhauling for small-cell {5G} networks,''
  \emph{IEEE Wireless Communications}, vol.~22, no.~5, pp. 83--89, Oct. 2015.

\bibitem{Zhang15a}
Z.~Zhang, X.~Wang, K.~Long, A.~Vasilakos, and L.~Hanzo, ``Large-scale
  {MIMO}-based wireless backhaul in {5G} networks,'' \emph{IEEE Wireless
  Communications}, vol.~22, no.~5, pp. 58--66, Oct. 2015.

\bibitem{Hong14a}
S.~Hong, J.~Brand, J.~Choi, M.~Jain, J.~Mehlman, S.~Katti, and P.~Levis,
  ``Applications of self-interference cancellation in {5G} and beyond,''
  \emph{IEEE Communications Magazine}, vol.~52, no.~2, pp. 114--121, Feb. 2014.

\bibitem{Ngo14}
H.~Q. Ngo, H.~Suraweera, M.~Matthaiou, and E.~Larsson, ``Multipair full-duplex
  relaying with massive arrays and linear processing,'' \emph{IEEE Journal on
  Selected Areas in Communications}, vol.~32, no.~9, pp. 1721--1737, Sep. 2014.

\bibitem{Nokia15a}
``Ten key rules of {5G} deployment,'' Nokia Solutions and Networks, white paper
  C401-01178-WP-201503-1-EN, 2015.

\bibitem{Kela15a}
P.~Kela, J.~Turkka, and M.~Costa, ``Borderless mobility in {5G} outdoor
  ultra-dense networks,'' \emph{IEEE Access}, vol.~3, pp. 1462--1476, Aug.
  2015.

\bibitem{Yang13}
H.~Yang and T.~Marzetta, ``Performance of conjugate and zero-forcing
  beamforming in large-scale antenna systems,'' \emph{IEEE Journal on Selected
  Areas in Communications}, vol.~31, no.~2, pp. 172--179, Feb. 2013.

\bibitem{Sahai13a}
A.~Sahai, S.~Diggavi, and A.~Sabharwal, ``On degrees-of-freedom of full-duplex
  uplink/downlink channel,'' in \emph{IEEE Information Theory Workshop (ITW)},
  Sep. 2013, pp. 1--5.

\bibitem{Kim15a}
K.~Kim, S.~W. Jeon, and D.~K. Kim, ``The feasibility of interference alignment
  for full-duplex {MIMO} cellular networks,'' \emph{IEEE Communications
  Letters}, vol.~19, no.~9, pp. 1500--1503, Sep. 2015.

\bibitem{Falaki10}
H.~Falaki, D.~Lymberopoulos, R.~Mahajan, S.~Kandula, and D.~Estrin, ``A first
  look at traffic on smartphones,'' in \emph{Proc. 10th ACM SIGCOMM Conference
  on Internet Measurement}, Nov. 2010, pp. 281--287.

\bibitem{Tulino04}
A.~Tulino and S.~Verd\'{u}, \emph{Random Matrix Theory and Wireless
  Communications}.\hskip 1em plus 0.5em minus 0.4em\relax Foundations and
  Trends in Communications and Information Theory, 2004.

\end{thebibliography}

\end{document}